\documentclass[9pt,twoside,letterpaper]{extarticle}

\usepackage[letterpaper,margin=0.75in]{geometry}
\usepackage[T1]{fontenc}

\usepackage{amsmath} 
\usepackage[numbers,sort&compress]{natbib} 
\usepackage[colorlinks = true]{hyperref} 
\usepackage{cancel, xcolor} 
\usepackage[normalem]{ulem} 
\usepackage{enumitem} 

\usepackage[OMLmathsfit]{isomath} 
\usepackage{textgreek} 
\usepackage{upgreek} 

\DeclareMathAlphabet{\matholdcal}{OMS}{cmsy}{m}{n} 

\usepackage[cal=dutchcal, scr=stixtwofancy]{mathalpha} 

\usepackage[bbgreekl]{mathbbol} 
\DeclareSymbolFontAlphabet{\mathbbm}{bbold}
\usepackage{amsfonts} 
\DeclareSymbolFontAlphabet{\mathbb}{AMSb}

\usepackage{amssymb}
\usepackage{stmaryrd} 

\renewcommand{\ss}[1]{\substack{#1}}
\newcommand{\bd}[1]{\boldsymbol{#1}}

\usepackage{tensor}
\newcommand{\idc}[1]{\indices{#1}}
\newcommand{\tld}[1]{\tilde{#1}}
\renewcommand{\Bar}[1]{\overline{#1}} 
\newcommand{\Tld}[1]{\widetilde{#1}} 

\usepackage{accents}

\newcommand{\lpr}{(}\newcommand{\rpr}{)}
\newcommand{\lbk}{\lbrack}\newcommand{\rbk}{\rbrack}
\newcommand{\lbe}{\lbrace}\newcommand{\rbe}{\rbrace}
\newcommand{\lag}{\langle}\newcommand{\rag}{\rangle}

\newcommand{\lrvt}[1]{|#1|} 
\newcommand{\lrpr}[1]{(#1)}
\newcommand{\lrbk}[1]{\lbk #1\rbk}
\newcommand{\lrbe}[1]{\lbe #1\rbe}
\newcommand{\lrag}[1]{\lag #1\rag}

\newcommand{\size}{\bigg} 

\newcommand{\Lpr}{\size(}\newcommand{\Rpr}{\size)}

\newcommand{\Lbk}{\size\lbrack}\newcommand{\Rbk}{\size\rbrack}

\newcommand{\Lag}{\size\langle}\newcommand{\Rag}{\size\rangle}

\newcommand{\Lrpr}[1]{\Lpr #1\Rpr}
\newcommand{\Lrbk}[1]{\Lbk #1\Rbk}

\newcommand{\Lrag}[1]{\Lag #1\Rag}

\newcommand{\oTr}{\operatorname{Tr}}
\newcommand{\oerf}{\operatorname{erf}}
\newcommand{\orelu}{\operatorname{relu}}

\newcommand{\ofp}{{\accentset{\scriptscriptstyle +}{\mathrm{FT}}}}

\newcommand{\mtx}[1]{\begin{matrix}#1\end{matrix}}
\newcommand{\pmx}[1]{\begin{pmatrix}#1\end{pmatrix}}
\newcommand{\bmx}[1]{\begin{bmatrix}#1\end{bmatrix}}
\newcommand{\smtx}[1]{\begin{smallmatrix}#1\end{smallmatrix}}
\newcommand{\spmx}[1]{\left\lpr\begin{smallmatrix}#1\end{smallmatrix}\right\rpr}
\newcommand{\sbmx}[1]{\left\lbk\begin{smallmatrix}#1\end{smallmatrix}\right\rbk}
\newcommand{\ee}{\mathrm{e}}
\newcommand{\ii}{\mathrm{i}}
\newcommand{\ppi}{{\text{\textpi}}}
\newcommand{\ddel}{\text{\reflectbox{6}}} 
\newcommand{\cchi}{{\text{\textchi}}} 
\newcommand{\ddelta}{{\text{\textdelta}}} 
\newcommand{\TTheta}{{\text{\textTheta}}} 
\newcommand{\inv}{\dashv}
\newcommand{\tsp}{\intercal}
\newcommand{\fl}{\sim} 

\newcommand{\dd}{\mathsf{d}}

\newcommand{\nam}[1]{\mathsfit{#1}}

\newcommand{\hid}[1]{#1} 
\newcommand{\Sm}[2]{\Lrbe{#1}_{#2}^{\Upsigma}} 
\newcommand{\sm}[2]{\lrbe{#1}_{#2}^{\Upsigma}}
\newcommand{\Pd}[2]{\Lrbe{#1}_{#2}^{\Uppi}} 
\newcommand{\pd}[2]{\lrbe{#1}_{#2}^{\Uppi}}
\newcommand{\Nt}[2]{\Lrbe{#1}_{#2}^{\smallint}} 
\newcommand{\nt}[2]{\lrbe{#1}_{#2}^{\smallint}}
\newcommand{\Ev}[1]{\Llrrbk{#1}} 
\newcommand{\ev}[1]{\llrrbk{#1}}

\newcommand{\Ep}[2]{\Lrag{#1}_{#2}} 
\newcommand{\ep}[2]{\lrag{#1}_{#2}}

\newcommand{\undoisomathgreek}{%
  \DeclareMathSymbol{\Gamma}{\mathalpha}{operators}{"00}%
  \DeclareMathSymbol{\Delta}{\mathalpha}{operators}{"01}%
  \DeclareMathSymbol{\Theta}{\mathalpha}{operators}{"02}%
  \DeclareMathSymbol{\Lambda}{\mathalpha}{operators}{"03}%
  \DeclareMathSymbol{\Xi}{\mathalpha}{operators}{"04}%
  \DeclareMathSymbol{\Pi}{\mathalpha}{operators}{"05}%
  \DeclareMathSymbol{\Sigma}{\mathalpha}{operators}{"06}%
  \DeclareMathSymbol{\Upsilon}{\mathalpha}{operators}{"07}%
  \DeclareMathSymbol{\Phi}{\mathalpha}{operators}{"08}%
  \DeclareMathSymbol{\Psi}{\mathalpha}{operators}{"09}%
  \DeclareMathSymbol{\Omega}{\mathalpha}{operators}{"0A}%
}
\undoisomathgreek

\renewcommand{\ee}{e}
\renewcommand{\ii}{i}
\renewcommand{\ppi}{\pi}
\renewcommand{\ddel}{\partial}
\renewcommand{\cchi}{\delta}
\renewcommand{\ddelta}{\delta}
\renewcommand{\TTheta}{\Theta}

\renewcommand{\inv}{{-1}}
\renewcommand{\tsp}{T}

\renewcommand{\dd}{d}

\renewcommand{\nam}[1]{#1}

\renewcommand{\hid}[1]{}
\renewcommand{\Sm}[2]{\sum_{#2}\Lrbk{#1}}
\renewcommand{\sm}[2]{\sum_{#2}#1}
\renewcommand{\Pd}[2]{\Lrbk{\prod_{#2}#1}}
\renewcommand{\pd}[2]{\Lrbk{\prod_{#2}#1}}
\renewcommand{\Nt}[2]{\int_{#2}\Lrbk{#1}}
\renewcommand{\nt}[2]{\int_{#2}#1}
\renewcommand{\Ev}[1]{\Lrpr{#1}}
\renewcommand{\ev}[1]{\lrpr{#1}}

\renewcommand{\eqref}[1]{Eq.~\ref{#1}}

\newcommand{\methref}[1]{Methods: \nameref{#1}}
\newcommand{\apperef}[1]{Appendix: \nameref{#1}}

\usepackage{tikz-cd}
\newcommand{\dbl}[1]{\mathchoice
  {\accentset{\scalebox{0.5}{$\displaystyle 2$}}{#1}}
  {\accentset{\scalebox{0.5}{$\textstyle 2$}}{#1}}
  {\accentset{\scalebox{0.5}{$\scriptstyle 2$}}{#1}}
  {\accentset{\scalebox{0.5}{$\scriptscriptstyle 2$}}{#1}}}
\newcommand{\crc}[1]{\mathchoice
  {\accentset{\scalebox{0.5}{$\displaystyle\circ$}}{#1}}
  {\accentset{\scalebox{0.5}{$\textstyle\circ$}}{#1}}
  {\accentset{\scalebox{0.5}{$\scriptstyle\circ$}}{#1}}
  {\accentset{\scalebox{0.5}{$\scriptscriptstyle\circ$}}{#1}}}
\newcommand{\sta}[1]{\mathchoice
  {\accentset{\scalebox{0.5}{$\displaystyle\star$}}{#1}}
  {\accentset{\scalebox{0.5}{$\textstyle\star$}}{#1}}
  {\accentset{\scalebox{0.5}{$\scriptstyle\star$}}{#1}}
  {\accentset{\scalebox{0.5}{$\scriptscriptstyle\star$}}{#1}}}
\newcommand{\oNor}{\mathscr{N}}
\newcommand{\oPR}{\mathrm{PR}}
\newcommand{\oOS}{\mathrm{OS}}
\newcommand{\oCS}{\mathrm{CS}}

\newcommand{\nni}{N}
\newcommand{\nnt}{{N_{\nam{t}}}}
\newcommand{\nnc}{{N_{\nam{c}}}}
\newcommand{\rhoin}{\rho_{\nam{f}}}
\newcommand{\mea}{{\nam{m}}}
\newcommand{\acov}{{\tld{\nam{C}}}}
\newcommand{\green}{{S}}
\newcommand{\XX}{{\mathsf{X}}} 

\usepackage{graphicx}
\usepackage{float}
\usepackage[font=small,labelfont=bf]{caption}
\usepackage{booktabs}
\usepackage{microtype}
\usepackage[english]{babel}
\usepackage{hyphenat}
\usepackage{balance} 

\newcommand{\dropcap}[1]{#1}

\makeatletter
\let\orig@section\section
\let\orig@subsection\subsection
\let\orig@subsubsection\subsubsection
\let\orig@paragraph\paragraph
\renewcommand{\section}{\@ifstar\named@section\orig@section}
\renewcommand{\subsection}{\@ifstar\named@subsection\orig@subsection}
\renewcommand{\subsubsection}{\@ifstar\named@subsubsection\orig@subsubsection}
\renewcommand{\paragraph}{\@ifstar\named@paragraph\orig@paragraph}
\newcommand{\named@section}[1]{\orig@section*{#1}\def\@currentlabelname{#1}}
\newcommand{\named@subsection}[1]{\orig@subsection*{#1}\def\@currentlabelname{#1}}
\newcommand{\named@subsubsection}[1]{\orig@subsubsection*{#1}\def\@currentlabelname{#1}}
\newcommand{\named@paragraph}[1]{\orig@paragraph*{#1}\def\@currentlabelname{#1}}
\makeatother

\graphicspath{ {figs/} }

\title{Random neural networks match observed dimensionality of neural population recordings and motivate stronger experimental tests}
\author{Zehui Zhao\textsuperscript{a} \and Michael J Pasek\textsuperscript{a, c} \and Ilya M Nemenman\textsuperscript{a, b, c}}
\date{\small \textsuperscript{a}Department of Physics, Emory University, Atlanta, GA 30322\\\textsuperscript{b}Department of Biology, Emory University, Atlanta, GA 30322\\\textsuperscript{c}Initiative for Theory and Modeling of Living Systems, Emory University, Atlanta, GA 30322}

\begin{document}

\twocolumn[
\begin{@twocolumnfalse}
\maketitle

\begin{abstract}
Randomly connected neural networks have long served as a theoretical tool for studying collective dynamics in neural populations, yet quantitative comparisons to experiments remain limited.
Recent technological advances have made it possible to resolve population-wide correlations across neurons, and minimal models such as random neural networks predict their generic structure. Whether the two agree quantitatively remains untested.
In this work, we examine whether a minimally structured random neural network can account for the low dimensionality of activity in neural population recordings by
building on recent developments in Dynamical Mean-Field Theory and incorporating two additional experimentally relevant features into the model: finite measurement time and variability across behavioral contexts.
We show that, when these factors are included, the dimensionality measured from large-scale recordings is consistent with the values predicted by random models. However, current recording durations make it difficult to use dimensionality to discriminate among connectivity structures. We further show that analytically predicted dimensionality varies non-monotonically with external input strength, and that the orientation similarity between neural manifolds recorded under different behavioral contexts can be more sensitive to network structure than dimensionality is. Together, these results provide quantitative guidance for experimental design to infer the connectivity structure underlying population activity.
\end{abstract}


\section*{Significance Statement}
Neural population activity in the brain is often low-dimensional: when many neurons are recorded simultaneously, their combined firing patterns vary in only a few coordinated ways rather than independently across neurons. This has been taken as evidence that the brain's wiring is specially organized to produce such simple, structured activity. We ask whether such low dimensionality could instead arise generically, even in networks with random connections without fine-tuning. Using analytical methods, we show that once realistic limits like finite recording time are accounted for, randomly connected networks produce activity whose dimensionality matches that observed in experiments. In other words, activity dimensionality measurements alone cannot distinguish random from structured wiring. We identify specific measurements, such as comparing activity geometry across behavioral conditions, that would provide stronger tests needed to resolve this question.

\noindent\textbf{Author contributions:} Z.Z.\ and I.M.N.\ designed the research; all authors performed the research and contributed to writing the manuscript.

\noindent\textbf{Competing interests:} The authors declare no competing interests.

\noindent\textbf{Correspondence:} Ilya M Nemenman, ilya.nemenman@emory.edu
\vspace{1em}
\end{@twocolumnfalse}
]

\dropcap{C}omplex systems with many interacting components are often well described by random network models, which capture essential behavior without fine-tuned parameter choices and have been used to identify what model structures are necessary to reproduce experimental observations~\cite{nemenman2025randomness}.
In neuroscience, such random models have been applied to large neural populations since the work of Sompolinsky and colleagues~\cite{sompolinsky1988Chaos}.
Yet, until recently, such theoretical work has aimed mostly at understanding the general principles of neural computation, rather than at  comparing to experiments.
Some attempts to connect models to experiments exist~\cite{rosenbaum2017Spatial, huang2019Circuit, sederberg2020Randomly, natale2020precise, tian2024Neuronal} (see additional references in \cite{nemenman2025randomness}), but, overall, theorists have not yet leveraged modern large-scale neural population recordings~\cite{saxena2019towards} to connect their models tightly to experimental data. 

One quantity that bridges random network models and modern recordings is the geometry of population activity, in particular its dimensionality~\cite{vyas2020computation,chung2021Neural,duncker2021dynamics,langdon2023unifying, perich2025neural}.
Dimensionality describes the system's effective number of degrees of freedom~\cite{gao2015simplicity,gao2017Theory,jazayeri2021interpreting,safaie2023preserved}.
Functionally, it reflects the number of independent variables encoded in the activity, including stimuli, tasks, and latent variables.
It therefore provides clues about the underlying computation~\cite{jazayeri2021interpreting}.
Experimentally, the dimensionality of population activity has been observed generally to be much lower than the full state space dimensionality $\nni$, that is, the number of recorded neurons~\cite{gao2017Theory, churchland2012Neural, mazzucato2016Stimuli, perich2018Neural, gallego2018Cortical, stringer2019Highdimensional, russo2020Neural, gallego2020Longterm, bartolo2020Dimensionality, cueva2020Lowdimensional, snyder2021Stable, altan2023Lowdimensional, manley2024Simultaneous, wang2025Geometry, oby2025Dynamical}.
This low dimensionality could reflect computation-specific organization in the circuit, modeled by introducing additional structure into the network~\cite{mastrogiuseppe2018Linking, clark2024Connectivity}.
We instead ask whether it can arise simply from recurrent interactions, modeled in the simplest case by a minimally structured random network.

Recently, Dynamical Mean-Field Theory (DMFT) methods have been developed to semi-analytically calculate the dimensionality of neural activity in a minimally structured random neural network in the infinite measurement time limit $T\to\infty$~\cite{clark2023Dimension}.
However, two crucial aspects of experimental recordings still need to be incorporated into the calculation for quantitative comparison with experiments.
First, experiments have finite durations, so the measured dimensionality reflects only the patterns visited during the recording window.
A system can therefore appear low-dimensional simply because the experiment is too short or samples too few conditions for the activity to explore its available state space~\cite{gao2017Theory}.
Second, neural systems receive inputs from other brain areas and from the external world~\cite{sauerbrei2020cortical,vahidi2024modeling,yang2021modelling,bachschmid2023interplay}, and these inputs change the measured dimensionality of population activity~\cite{bartolo2020Dimensionality}.
Existing analyses of dimensionality in random networks treat the network as autonomous~\cite{clark2023Dimension}.
In this work, we develop analytical approaches to address both issues.

We calculate the dimensionality of population activity in the minimally structured network and show that it is quantitatively consistent with primate motor cortex data~\citep{gao2017Theory}.
However, in the regime set by current recording durations, the predicted dimensionality is low and insensitive to many network parameters, so this agreement may simply reflect the limited measurement window.
Even though the agreement is better than for independent neurons~\citep{gao2017Theory}, dimensionality alone is insufficient to identify the correct model of population activity.
We therefore propose four additional measurements that probe network structure beyond dimensionality, and calculate predictions for these measurements in the random network model.
First, the dimensionality of activity fluctuations varies non-monotonically with external input strength, a qualitative signature easier to detect than the absolute number of dimensions.
Second, the cosine similarity between time-averaged activity patterns under two behavioral contexts depends only weakly on input strength and saturates, so it is not a strong probe of emergent population coding.
Third, the orientation of the high-dimensional fluctuation ellipsoid changes much more rapidly with input strength than this cosine similarity does, so two behavioral contexts can yield similar mean responses while differing substantially in their fluctuation structure.
Fourth, the dimensionality of mean activity patterns across many contexts grows in a predictable way with the number of contexts sampled, providing a baseline against which experimental data can be compared.
For each quantity, we derive quantitative predictions and identify what experimental outcomes would point to additional organization beyond the minimally structured baseline.

\section*{Results}

\subsection*{Model setup}\phantomsection\label{sec:model}

To model a population of neurons with minimally structured connectivity under behavioral context, we adopt the standard random recurrent neural network of Sompolinsky and colleagues~\cite{sompolinsky1988Chaos, clark2023Dimension, engelken2023Lyapunov} and add external inputs to represent the behavioral context.
Both the coupling matrix and the external inputs are random and minimally structured.
Concretely, for neurons indexed by $i=1,\cdots,\nni$, the net current $h_i\ev{t}$ to neuron $i$ at time $t$ evolves according to
\begin{equation}
  \label{eq:ode}
  h_{i}\ev{t}+\ddel_th_{i}\ev{t} = \sm{J_{ij}r_{j}\ev{t}}{j}+f_i,
\end{equation}
where $J$ is the coupling matrix, $r_j\ev{t}$ is the firing rate of neuron $j$, and $f_i$ is the external input to neuron $i$.
The rate $r_i$ and current $h_i$ of every neuron are related by a nonlinearity $\phi$ as $r_{i}=\phi\ev{h_{i}}$.
We also refer to $r$ and $h$ as the activation and preactivation, respectively.
We choose $\phi$ to be odd so that the network can be symmetric under $h\to -h$ and the means of the rate and current are $0$.
The qualitative behavior of the network is expected to be robust to this choice, as verified in similar contexts numerically~\cite{rajan2010Stimulusdependent, mastrogiuseppe2017Intrinsicallygenerated}.
Specifically, we choose $\phi\ev{h}=\oerf\ev{\sqrt{\ppi}h/2}$, so that the rate's variance under gaussian currents have closed-form expressions~\cite{vanmeegen2021Microscopic}.
Overall, on the l.h.s.\ of \eqref{eq:ode}, we have the continuous time equilibration of the net current~\cite{dayan2005Theoretical}, and the time here is measured in units of the single-neuron membrane constant, not to be confused with the autocorrelation time that emerges in the presence of interactions.
The two terms on the r.h.s.\ of \eqref{eq:ode} represent the additional input to the current at time $t$.
The first term models interactions between neurons, and the second models the external inputs.
While in this paper we work with specific choices for the nonlinearity $\phi$, the statistics of the coupling matrix $J$, and external input $f$, many of the calculations can be reproduced for other choices as well.

In the interaction term, the input from the $j$-th neuron is proportional to its firing rate $r_j$, and the proportionality constant is the coupling matrix element $J_{ij}$.
The proportionality constants in the coupling matrix $J$ can be thought of as the signed effective synaptic weights, but we do not constrain them to respect Dale's law, since in general the effects from inhibitory neurons can be both inhibitory or excitatory through disinhibition mediated by recurrent pathways, and similarly for excitatory neurons. 
Since we want the interactions to be generic without any fine-tuning or constraints, we adopt the standard choice~\cite{sompolinsky1988Chaos, clark2023Dimension} of sampling $J$ from the minimally structured i.i.d.\ zero-mean Gaussian distribution $J_{ij}\fl\oNor\ev{0,g^2/\nni}$.
We refer to $g$ as the gain parameter.
A nonzero mean would add a rank-one component to the coupling matrix, and since we want our coupling matrix to be minimally structured for calculating the least fine-tuned behavior, we choose the mean to be $0$.
The $1/\nni$ scaling in the variance of coupling constants $J_{ij}$ can be interpreted as each neuron maintaining a fixed number of connections with fixed coupling strengths as the neuron number $\nni$ is increased, which ensures the interaction term does not blow up with the neuron number $\nni$.
The network's behavior depends only on the variance of $J_{ij}$~\cite{mastrogiuseppe2017Intrinsicallygenerated}, and the Gaussian choice simplifies calculations.

To account for the fact that the dynamics of a neural population depends on the behavioral context, including both the internal state of the animal and its external stimuli~\cite{churchland2010Stimulus}, we assume that the neural population dynamics is modulated by explicit inputs from its upstream populations.
We choose the external inputs to be time-independent, which can be seen as the adiabatic approximation to slowly-varying external inputs.
We refer to each time-independent external input as one \emph{behavioral context}, but the term can be broadly understood to refer to any period of approximately constant inputs to the dynamics.
Since we are interested in the behavior of neural populations under generic conditions, we again sample the external inputs from the minimally structured i.i.d.\ zero-mean Gaussian distribution $f_i\fl\oNor\ev{0,I^2}$, where the standard deviation $I$ represents the external input strength.
This choice of external time-independent inputs is also referred to as quenched noise~\cite{schuecker2018Optimal}.
As before, the Gaussian and zero-mean choices simplify calculations (see Appendix) and do not affect qualitative behavior.
A nonzero input mean would prevent the rate's variance from having a closed form~\cite{vanmeegen2021Microscopic, dick2024Linking}.

\subsection*{Treating external inputs}

In the absence of external inputs, \eqref{eq:ode} describes the autonomous dynamics of the network, which has been thoroughly studied using DMFT~\cite{sompolinsky1988Chaos, clark2023Dimension, engelken2023Lyapunov}.
In the limit of a large number of neurons $\nni\to\infty$, the autonomous system is self-averaging and stationary to the leading order after the initial transient~\cite{sompolinsky1988Chaos}.
This means the population autocovariance is independent of the realization of the coupling matrix $J$ and of the absolute time $t$, and can be written as a function of the lag $\tau$ only,
\begin{equation}
  C_\tau = \frac{1}{\nni}\sm{r_{i}\ev{t}r_{i}\ev{t+\tau}}{i}.
\end{equation}
The autocovariance $C_\tau$ is the order parameter that characterizes the dynamical regime of the system: for $g<1$, the quiescent state has $C_{\tau=0}=0$; for $g>1$, chaotic activity has $C_0>0$ and $C_\tau$ decays to $C_\infty=0$ as the lag $\tau$ increases.
Equivalently, the network has a positive maximum Lyapunov exponent when $g>1$~\cite{sompolinsky1988Chaos, engelken2023Lyapunov}.
The transition to chaos as the gain parameter $g$ increases is found for a broad range of coupling distributions beyond the i.i.d.\ Gaussian $J$ used here~\cite{aljadeff2015Transition, aljadeff2016Lowdimensional, 
  marti2018Correlations, shao2024Identifying, 
  kadmon2015Transition, mastrogiuseppe2018Linking, herbert2022Impact, shao2023Relating, dick2024Linking, 
  rajan2006Eigenvalue, landau2018Coherent, hayakawa2020Spontaneous, landau2021Macroscopic, harris2023Effect, 
  stern2014Dynamics, stern2023Reservoir, 
  khajeh2022Sparse, 
  ostojic2014Two, krishnamurthy2022Theory}.

For a broad range of external inputs, such as sinusoids, time-independent constants, and white-noise, the network activity can be decomposed into two components, one being the response elicited by the external input, and the other being the modified autonomous activity on top of the response~\cite{rajan2010Stimulusdependent, engelken2022Input, molgedey1992Suppressing, schuecker2018Optimal}.
We refer to the two components as the \emph{ordered response} $\bar{r}$ and the potentially quiescent \emph{temporal chaos} $\tld{r}$, respectively.
Throughout, we mark quantities associated with ordered responses with a bar and quantities associated with temporal chaos with a tilde. For three-index quantities, we separate indices with commas (e.g., $h_{a,i,t}$); for two-index pairs like $h_{it}$ we follow the conventional concatenated notation.
The ordered response often resembles the form of the input, and for time-independent external inputs, the ordered response is also time-independent, with
\begin{equation}
  \label{eq:r-decomp}
  \bar{r}_i = \ep{r_i\ev{t}}{t},\quad \tld{r}_i\ev{t} = r_i\ev{t}-\bar{r}_i,
\end{equation}
where $\ep{\cdots}{x}$ denotes averaging over $x$.
The dynamical regime is determined by the residual fluctuation $\tld{r}$, so the order parameter in the driven network is the autocovariance of $\tld{r}$:
\begin{equation}
  \label{eq:op1-tld}
  \tld{C}_\tau = \frac{1}{\nni}\sm{\tld{r}_{i}\ev{t}\tld{r}_{i}\ev{t+\tau}}{i}.
\end{equation}
We separately describe the per-neuron variance of $\bar{r}$ as (recall that mean activity of each neuron is zero)
\begin{equation}
  \label{eq:op1-bar}
  \bar{C} = \frac{1}{\nni}\sm{\bar{r}_{i}^2}{i}.
\end{equation}
Geometrically, the ordered response $\bar{r}$ sets the center of the activity cloud, while the temporal chaos $\tld{r}$ determines the shape of the cloud around that center.
Approximating this cloud by an ellipsoid, $\tld{C}_{\tau=0}$ describes its size, and $\bar{C}$ describes the squared distance of its center from the origin.
Figure~\ref{fig:tld}A shows this picture in a schematic $2$-dimensional projection, with temporal chaos shown by the blue trajectory and the ordered response shown by the yellow cross.
In what follows, we use fluctuations of $\tld{C}$ across realizations to compute the dimensionality of this ellipsoid, quantified by the participation ratio (PR).

\subsection*{Modeling finite measurement time}

Neural recordings have a finite measurement time $T$, whereas DMFT gives quantities in the long-time limit $T\to\infty$.
To model the effect of finite $T$ on each statistic of the activity, we define the weight function (with $\TTheta$ denoting the Heaviside step function)
\begin{equation}
  \label{eq:weight-1}
  w_{T t_\mea}\ev{t} = \frac{1}{T}\TTheta\ev{t-\lrpr{t_\mea-T/2}}\TTheta\ev{\lrpr{t_\mea+T/2}-t}
\end{equation}
for a time window of measurement centered at $t_\mea$ of length $T$, so that the weight function takes value $1/T$ for $t_\mea-T/2<t<t_\mea+T/2$ and $0$ otherwise.
This allows us to express the mean and covariance of temporal chaos measured over the corresponding time window as
\begin{equation}
  \label{eq:finite-stat}
  \begin{aligned}
    &\bar{r}_{T t_\mea i} = \int \dd t\, w_{T t_\mea}\ev{t} r_{i}\ev{t},\\
    &\tld{\Sigma}_{T t_\mea ij} = \int\dd t \,w_{T t_\mea}\ev{t} \lrpr{\tld{r}_{i}\ev{t} - \bar{r}_{T t_\mea i}} \lrpr{\tld{r}_{j}\ev{t} - \bar{r}_{T t_\mea j}},
  \end{aligned}
\end{equation}
respectively.
Within the integrals, the neural activity $r$ is described by the long-time statistics given by DMFT.
So by averaging different finite-time statistics over the window location $t_\mea$, we can relate them to their long-time values.
Since the external inputs are time-independent or slowly varying, predictions in our model assume stationary dynamics or a fixed behavioral context.
In experiments with non-stationary dynamics or multiple behavioral contexts, $T$ in \eqref{eq:weight-1} and \eqref{eq:finite-stat} is the duration over which the dynamics is approximately stationary.

By the central limit theorem, the error in the finite-time ordered response has a variance scaling as $\sim \tau_\acov/T$, where $\tau_\acov$ is the width of the autocovariance function $\tld{C}_\tau$ (i.e., the autocorrelation time. Details can be found in \apperef{sec:error-over-time}).
In the cortex, the autocorrelation time is often on the order of $100$ ms~\cite{murray2014Hierarchy}, while a typical behavior such as a reaching movement lasts on the order of $1$ s~\cite{gao2017Theory}.
So in the experimentally relevant regime, $T/\tau_\acov \approx 10$, such that the ordered response is relatively well measured.
This allows   us to simplify the finite-time covariance $\tld{\Sigma}_{T t_\mea}$ in \eqref{eq:finite-stat} into
\begin{equation}
  \label{eq:finite-cov}
  \tld{\Sigma}_{T t_\mea ij} = \int\dd t \,w_{T t_\mea}\ev{t} \tld{r}_{i}\ev{t}\tld{r}_{j}\ev{t}
\end{equation}
by approximating the finite-time ordered response $\bar{r}_{T t_\mea}$ with the true long-time ordered response $\bar{r}$.
This approximate expression simplifies later calculations on $\tld{\Sigma}_{T t_\mea}$.

The autocorrelation time $\tau_\acov$ controls how the dimensionality measured over a window of length $T$ grows with $T$~\cite{gao2017Theory}.
For independent neurons where the activity of each neuron is described by some autocorrelation time $\tau_\acov$, the linear dimensionality of the activity is approximately constant in $T$ when $T\ll\tau_\acov$, and grows linearly with $T$ with growth rate $\sim 1/\tau_\acov$ when $T\gtrsim\tau_\acov$. Intuitively, every $\tau_\acov$, the trajectory yields one statistically independent sample, and in a high-dimensional state space each new sample lies along a new direction.
The width of $\tld{C}_\tau$ admits several conventional definitions; we choose $\tau_\acov$ so that the dimensionality of independent neurons grows at rate exactly $1/\tau_\acov$, yielding a unit rate of dimensionality increase when measured in rescaled time $T/\tau_\acov$.
Concretely, $\tau_\acov$ is defined as
\begin{equation}
  \label{eq:tau-acov}
  \tau_\acov = \int\dd \tau \Lrpr{\frac{\tld{C}_\tau}{\tld{C}_{\tau=0}}}^2.
\end{equation}

\begin{table}[t!]
  \centering
  \caption{Table of variables}
  \begin{tabular}{ll}
    variable & description\\
    \midrule
    $\nni$ & neuron number\\
    $T$ & measurement time\\
    $\nnc$ & number of behavioral contexts\\
    $g$ & gain parameter\\
    $I$ & external input strength\\
    $J$ & coupling matrix\\
    $f$ & external input\\
    $h$ & input current to neuron\\
    $\phi$ & nonlinearity between $h$ and $r$\\
    $r$ & firing rate of neuron\\
    $\bar{r}$ & ordered response\\
    $\tld{r}$ & temporal chaos\\
    $\bar{C}$ & variance of ordered response\\
    $\tld{C}_\tau$ & autocovariance of temporal chaos\\
    $\tau_\acov$ & width of autocovariance\\
    $\Tld{\oPR}_T$ & finite-time dimensionality of temporal chaos\\
    $\bar{C}_{12}$ & two-replica overlap of ordered response\\
    $\tld{C}_{12\tau}$ & two-replica overlap of temporal chaos\\
    $\oCS$ & similarity between ordered responses\\
    $\oOS$ & similarity between temporal chaos\\
    $\Bar{\oPR}_\nnc$ & finite-context-number dimensionality \\
    & of ordered response\\
    $C^{\nam{h}}_\tau$ & preactivation autocovariance\\
    \bottomrule
  \end{tabular}
\end{table}

\subsection*{Geometry of activity during a single behavioral context}\phantomsection\label{sec:tld}

We first describe the geometry of activity when the neural population is under a fixed behavioral context.
In the model in \eqref{eq:ode}, this corresponds to the activity under a single realization of the coupling matrix $J$ and external input $f$.

\begin{figure}[bt!]
  \centering
  \includegraphics{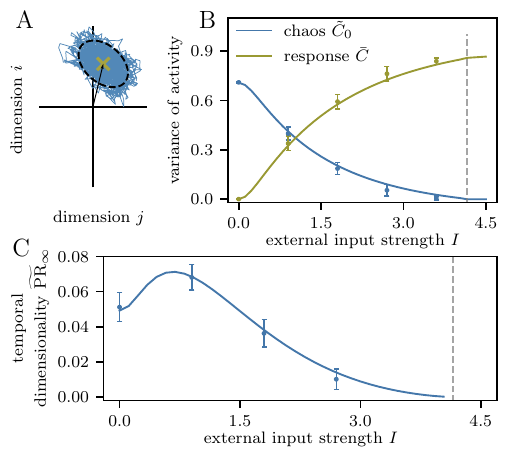}
  \caption{Statistics of activity under a single behavioral context.
    A: Schematic: blue curves represent the temporal chaos, approximated as an ellipsoid in state space; the yellow cross marks the ordered response at the ellipsoid's center.
    B: The variance of temporal chaos $\tld{C}_0$ decreases with external input strength $I$, while the variance of the ordered response $\bar{C}$ increases with $I$, as indicated by the legend.
    The ordered-response variance $\bar{C}$ shows a cusp at the transition from chaotic to ordered dynamics, where $\tld{C}_0 = 0$, marked by the vertical dashed line.
    For $g=3$, the transition happens around $I\approx 4.15$.
    Markers with error bars represent simulation results (mean and standard deviation over realizations) with $\nni=800$.
    C: The long-time dimensionality $\Tld{\oPR}_\infty$ is non-monotonic in external input strength.
  }
  \label{fig:tld}
\end{figure}

For a broad range of external inputs, the size of temporal chaos, as measured by $\tld{C}_\tau$, is known to decrease with input strength~\cite{rajan2010Stimulusdependent, engelken2022Input, molgedey1992Suppressing, schuecker2018Optimal}.
Temporal chaos eventually vanishes when the external input strength reaches some critical value, and the network transitions back to ordered dynamics.
Intuitively, as long as the external input drives the network to regions of the nonlinearity with lower gain $\phi'$, the network will be less unstable (lower Lyapunov exponent) and chaos will be weaker.
We first confirm this in our model.
Since the external inputs we consider are time-independent, the size of temporal chaos is simply described by its variance $\tld{C}_0$, and the point of transition is marked by $\tld{C}_0=0$.
We compute $\bar{C}$ and $\tld{C}_0$ semi-analytically from the single-replica DMFT equations, \eqref{eq:1-repl-dmft-bar} and \eqref{eq:1-repl-dmft-tld}, derived in \methref{sec:dmft1-sp} (see also~\cite{sompolinsky1988Chaos, clark2023Dimension}).
As shown in Figure~\ref{fig:tld}B, under a fixed gain parameter $g=3>1$, as the external input strength $I$ increases, temporal chaos is gradually suppressed, reflected by the monotonic decrease in the variance of temporal chaos $\tld{C}_0$.
Eventually, the network transitions to ordered dynamics, marked by $\tld{C}_0=0$ at the dashed line (around $I\approx 4.15$ for $g = 3$).

Simultaneously, $\bar{C}$ monotonically increases with external input strength $I$, as shown in the same panel.
As the external input strength increases, the growth of $\bar{C}$ with $I$ slows, and slows further at the transition.
Intuitively, the initial slowdown is due to the network being driven to regions of the nonlinearity with lower slope $\phi'$, so the marginal expansion of the variance is more restricted.
The second slowdown reflects that the ordered response is a time average after the nonlinearity, so shrinking temporal chaos contributes to increasing the magnitude of $\bar{C}$ before the transition; once temporal chaos vanishes, this contribution is lost.

To verify our calculations, we simulate numerically a network described by \eqref{eq:ode} and compute the variances $\bar{C}$ and $\tld{C}_0$ from the simulated network activity, shown in Fig.~\ref{fig:tld}B as markers with error bars.
Error bars are standard deviations across realizations, indicating the level of self-averaging at biologically realistic neuron numbers.
Studies mapping neurons and their connections reveal that the network distance over which connection probability starts to decrease is about $100-1000$ neurons~\cite{holmgren2003Pyramidal, herculano-houzel2013Distribution}, and we use $\nni = 800$.
The simulation values agree with the semi-analytic $\nni\to\infty$ predictions, confirming that finite-size corrections remain small at this $\nni$.

In experiments, $\bar{r}$ corresponds to the time-averaged activity under a fixed behavioral context (relative to baseline), and $\tld{r}$ to its temporal fluctuations, with $\bar{C}$ and $\tld{C}_0$ their population variances.
The qualitative behavior of both variances over external input strength $I$ that we describe here should hold for any system where the gain is maximal at the resting state.
Measuring the two variances in experiments where the input strength from other regions is known to increase thus reveals where single neurons sit on the nonlinearity.

\subsection*{Long-time dimensionality of activity is non-\\monotonic in external input strength}\phantomsection\label{sec:tld-long}

We now turn to the dimensionality of temporal chaos and show that it varies non-monotonically with external input strength.
The dimensionality can either be quantified linearly or nonlinearly, but it has been shown recently that the two are likely to be similar for random networks~\cite{clark2023Dimension, engelken2023Lyapunov}.
We choose the linear dimensionality for two reasons.
First, it is semi-analytically calculable for all values of the gain parameter $g$, whereas the nonlinear Lyapunov dimensionality is calculable only for limiting values of $g$~\cite{engelken2023Lyapunov}.
Second, the linear dimensionality is more straightforward to generalize to finite measurement times $T$.
Specifically, we use the participation ratio (PR) of the principal component variances of temporal chaos~\cite{clark2023Dimension}, defined as
\begin{equation}
  \label{eq:pr-tld}
  \Tld{\oPR}\ev{\tld{\Sigma}_\infty} =\frac{\ep{\tld{\lambda}_i}{i}^2}{\ep{\tld{\lambda}_i^2}{i}}= \frac{\lrpr{\oTr\ev{\tld{\Sigma}_{\infty}}}^2}{\nni\oTr\ev{\tld{\Sigma}\idc{_\infty^2}}},
\end{equation}
where $\tld{\lambda}_i$ is the $i$-th eigenvalue of the covariance matrix $\tld{\Sigma}_{\infty}$.
This matrix is related to \eqref{eq:finite-stat} as its long-measurement-time limit $T\to\infty$:
\begin{equation}
  \label{eq:cov-tld}
  \tld{\Sigma}_{\infty ij} = \ep{\tld{r}_{i}\ev{t}\tld{r}_{j}\ev{t}}{t},
\end{equation}
and $\ep{\cdot}{i}$ again denotes expectation over $i$.
Intuitively, the PR measures the roundness of the ellipsoid of temporal chaos (Figure~\ref{fig:tld}A): it counts how many directions of activity carry comparable variance, normalized by $\nni$, and is independent of the overall variance.
The PR therefore takes values between $1/\nni$ and $1$.

In the numerator of \eqref{eq:pr-tld}, the trace of $\tld{\Sigma}$ is $\nni\tld{C}_0$ according to \eqref{eq:op1-tld}, which we already calculated in Figure~\ref{fig:tld}B.
We therefore only need to calculate the trace of its square, $\tilde{\Sigma}^2$, in the denominator.
Conveniently, it is related to the lower-order (subleading) fluctuation over time of the autocovariance $\tld{C}$ of temporal chaos:
\begin{equation}
  \label{eq:cov-tld-sq}
  \begin{aligned}
    \oTr\ev{\tld{\Sigma}\idc{_\infty^2}}
    &= \sm{\ep{\tld{r}_{it_1}\tld{r}_{it_2}\tld{r}_{jt_1}\tld{r}_{jt_2}}{t_1t_2}}{ij}
    = \nni^2\ep{\tld{C}_{t_1t_2}^2}{t_1t_2}\\
    &= \nni^2\ep{\tld{C}_{t,t+\infty}^2}{t},
  \end{aligned}
\end{equation}
where the time dependence of temporal chaos $\tld{r}$ is moved to the subscript, and $\tld{C}_{t_1t_2}$ is the time-dependent autocovariance, including both the stationary component $\tld{C}_{t_2-t_1}$ in the leading order and the subleading  $\sim \pm 1/\sqrt{\nni}$ time-dependent fluctuation.
The last line in \eqref{eq:cov-tld-sq} follows from the fact that since $\tld{C}_{t,t+\tau}$ is exponentially suppressed with $\tau$ in the chaotic regime, its value quickly converges in $\tau$ to $\tld{C}_{t,t+\infty}$, and it takes a different value only over a finite interval $\tau\lesssim \tau_\acov$ in the infinite domain over which $t$ is averaged.
So combining the numerator and the denominator, the dimensionality over long times $T\to\infty$ is
\begin{equation}
  \label{eq:pr-tld-result}
  \Tld{\oPR}_\infty = \frac{\tld{C}_0^2}{\nni\ep{\tld{C}_{t,t+\infty}^2}{t}}.
\end{equation}
Since the fluctuation in the autocovariance $\tld{C}$ is subleading, the denominator is $\sim_{\hid{\nni}} 1$ when $\nni$ is large, so $\Tld{\oPR}_\infty\sim_{\hid{\nni}} 1$.
Subleading fluctuations of $\tld{C}_{t,t+\infty}$ must be retained because they give the leading nonzero contribution to the PR denominator.
Subleading corrections to $\tld{C}_0$, however, can be neglected, because the numerator has a nonzero leading-order saddle-point value.
This means that in our minimally structured network, for a fixed gain parameter $g$ and external input strength $I$, the number of dimensions explored by temporal chaos over long times is a fixed fraction of the total neuron number, as in the autonomous case~\cite{clark2023Dimension}.
From \eqref{eq:cov-tld-sq}, this also implies that the cross-covariance between different neurons $\tilde{\Sigma}_{ij}$ is on the order of $1/\sqrt{\nni}$.

We compute the variance $\ep{\tld{C}_{t,t+\infty}^2}{t}$ from fluctuations around the single-replica saddle point, under a self-averaging assumption that may break down near the transition to ordered dynamics, as detailed in \methref{sec:dmft1-fluct}.
The resulting values of the long-time temporal chaos $\Tld{\oPR}_\infty$ as a function of the external input strength $I$ are shown in Figure~\ref{fig:tld}C.
Unlike the variances in Figure~\ref{fig:tld}B, the dimensionality varies non-monotonically as the external input strength $I$ increases, first increasing to around $50\%$ above the value at $I=0$, then decreasing towards $1/\nni$ and becoming undefined as the network transitions from chaos to ordered dynamics, marked in the Figure by the dashed line.
We find that the non-monotonic behavior persists over orders of magnitude of the gain parameter $g$
(see details in \apperef{sec:g-indp}).

Intuitively, the non-monotonicity reflects a change in the dominant mechanism shaping temporal chaos.
At low external input strength, the fluctuation-dissipation relation implies that the ordered response shifts the trajectory along directions where temporal chaos has large variance (as shown numerically in \apperef{sec:fluct-dissp}), so these high-variance directions saturate first and the ellipsoid becomes rounder.
As the input strength increases further, the external input and ordered response push some neurons into low-gain regions of the nonlinearity throughout the trajectory, lowering the effective recurrent gain and weakening temporal chaos.
The maximum in $\Tld{\oPR}_\infty$ marks the crossover between these two regimes.
Near the chaos-to-order transition, self-averaging breaks down and the DMFT predictions should be interpreted with caution.

Although interesting theoretically, we note that the long-time dimensionality is hard to measure experimentally for most behaviors because the behavioral context must remain constant over a significant period of time.

\subsection*{Dimensionality increases slower over measurement time for fewer neurons}\phantomsection\label{sec:tld-short}

Having found the dimensionality in the long measurement time limit $T\to\infty$, we now ask how the dimensionality under a fixed behavioral context depends on the measurement time $T$.
We calculate the dimensionality $\Tld{\oPR}_T$ as a function of $T$ for generically interacting neurons under generic behavioral contexts.
For experimentally relevant times $T/\tau_\acov \approx 10$ and biologically realistic $\nni$, the system is in the regime $T \lesssim \nni$.
In the large-$\nni$ limit $T \ll \nni$, dimensionality grows linearly with $T$.
For realistic $\nni$, correlations between neurons slow this growth to sublinear.
For longer measurement times $T\gg \nni$, the dimensionality slowly saturates as $\nni/T$.

\begin{figure}[t!]
  \centering
  \includegraphics{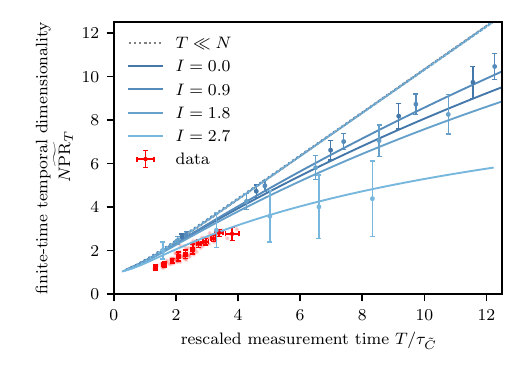}
  \caption{Dependence of the finite-time dimensionality $\Tld{\oPR}_T$ on measurement time $T$ and its comparison to data.
    The large-network limit $T\ll\nni$ is shown in dashed lines, where the dimensionality increases linearly, same as for independent neurons.
    For realistic values of the number of neurons  $\nni$ and autocovariance time $\tau_{\tilde{C}}$, the dimensionality grows sublinearly with rates dependent on the external input strength $I$.
    Compared to data measured in monkeys' PMd and M1 areas while the monkeys perform an eight-direction center-out delayed reach task, the dimensionality can be as low as the experimental values (binned means and standard deviations from data extracted and rescaled from \citep{gao2017Theory}, see \methref{sec:numerics}).
    $N=800$, $g=3$. }
  \label{fig:finite-tld}
\end{figure}

Using the finite-time covariance of temporal chaos in \eqref{eq:finite-stat} and the expression for the long-time dimensionality $\Tld{\oPR}_\infty$ in \eqref{eq:pr-tld}, we show in \methref{sec:finite} that the finite-time linear dimensionality is 
\begin{equation}
  \label{eq:finite-pr-tld-result}
  \begin{aligned}
    \Tld{\oPR}_T& = \ep{\oPR\ev{\tld{\Sigma}_{T t_\mea }}}{t_\mea}\\
    &= \frac{\tld{C}_0^2}{\nni\ep{\tld{C}_{t,t+\infty}^2}{t} +\nni\int\dd\tau\frac{\orelu\ev{1 - \lrvt{\tau}/T}}{T}\tld{C}_\tau^2},
  \end{aligned}
\end{equation}
where $\orelu$ is the rectified linear function $\orelu\ev{x}\equiv$ $\mathrm{max}\ev{0,x}$.
The derivation and approximation are given in \methref{sec:finite} and confirmed against simulations in Figure~\ref{fig:finite-tld}.
Compared to the long-time dimensionality $\Tld{\oPR}_\infty$ in \eqref{eq:pr-tld-result}, the finite-time dimensionality $\Tld{\oPR}_T$ has an additional non-negative integral correction in the denominator, thus reducing the dimensionality.
We can therefore expect the finite-time dimensionality to also increase with $g$ and vary non-monotonically with $I$, as for its long-time limit $\Tld{\oPR}_\infty$ in Figure~\ref{fig:tld}C.
This expression also shows that the absolute number of dimensions $\nni\Tld{\oPR}_T$ increases with $\nni$, where the increase is in general nonlinear.
Crucially, we cannot vary $g$, $I$, and $\nni$ to fit low dimensionalities, since random network models have the implicit assumption that the network should be relatively self-averaging to be described by the analysis.
Decreasing $\nni$ explicitly increases finite-size fluctuation, and so does moving the network closer to transition by increasing $I$ or decreasing $g$.

Inside the integral correction, $\tld{C}_\tau$ decays over width $\tau_\acov$~\cite{sompolinsky1988Chaos}, while the $\orelu$ factor has width $\sim T$.
When $T\ll\tau_\acov$, the integral selects $\tld{C}_0^2$, giving $\Tld{\oPR}_T=1/\nni$, or one dimension.
When $T\gg\tau_\acov$, the integral is $\tld{C}_0^2\tau_\acov/T$ by \eqref{eq:tau-acov}, yielding
\begin{equation}
  \label{eq:finite-pr-tld-result-approx}
  \Tld{\oPR}_T \stackrel{T\gg\tau_\acov}{\approx} \frac{\tld{C}_0^2}{\nni\ep{\tld{C}_{t,t+\infty}^2}{t} +\nni\frac{\tau_\acov}{T}\tld{C}_0^2}.
\end{equation}
The correction scales as $\nni\tau_\acov/T$, so reaching the long-time limit requires $T/\tau_\acov\gg\nni$.
This extra factor of $\nni$, compared with the $\sim\tau_\acov/T$ error in the finite-time ordered response discussed in \apperef{sec:error-over-time}, reflects the number of samples required to resolve the covariance spectrum.
Measured dimensionality in experiments is therefore unlikely to reach the long-time limit within a fixed behavioral context.

When the number of neurons $\nni$ is large enough such that $1 \ll T/ \tau_\acov\ll\nni$, the dimensionality grows linearly as $\nni\Tld{\oPR}_T = T/\tau_\acov$ as if the neurons were independent.
But for smaller $\nni$ not satisfying this separation between scales, the dimensionality would directly enter the saturating regime $T/\tau_\acov\ll\nni$ after the $T$-independent regime $T\gg\tau_\acov$, resulting in sublinear growth.
This suggests that under generic interactions, the dimensionality measured for short behaviors will only reflect correlations if the network size is not too large.
As argued above, the cortical parameters ($\tau_\acov \sim 100$ ms, $T \sim 1$ s, $\nni \sim 100$--$1000$) satisfy $T/\tau_\acov \lesssim \nni$.

Figure~\ref{fig:finite-tld} shows the semi-analytic solutions to \eqref{eq:finite-pr-tld-result} for realistic values of the model parameters $\nni$, $T$ and $\tau_\acov$ ($g$).
Simulations agree with the theory, with fluctuations increasing for longer $T$ and stronger external input.
Near the transition, self-averaging weakens, so finite-size deviations become larger.

Given the dimensionality predicted by the minimally structured random network, we are interested in whether experimental data constrain the validity of the model or its effective parameters.
Ref.~\citep{gao2017Theory} reports PR dimensionality from preprocessed data recorded from monkeys' PMd and M1 areas while the monkeys perform an eight-direction center-out delayed reach task.
The extracted and rescaled data points are summarized by binned means and standard deviations along with our predictions in Figure~\ref{fig:finite-tld}, and the extraction, rescaling, and binning procedure is described in \methref{sec:numerics}.
At shorter measurement times $T\lesssim\tau_\acov$, the model does not agree with experimental data.
This is unsurprising, because we expect short-timescale features of the trajectory to depend on physiological details of the model, such as the stereotypical way the firing rate rises and drops in a single neuron, when and how synaptic transmission occurs, etc.
This could be incorporated into the model by adjusting the single-neuron or transmission details, such as the shape of the nonlinearity, form of the causal operator (l.h.s.\ of \eqref{eq:ode}), or potential latency in the interaction, but we have not fine-tuned such details in our model.
Longer measurement times $T\gtrsim\tau_\acov$, on the other hand, are more likely to reflect collective computation in the system, such as how activity spreads across the network, controlled by structure in the coupling matrix.
At these longer measurement times, our minimal random network produces dimensionalities as low as the experimental values.
We note, however, that large input strengths near the chaos-to-order transition are where finite-size corrections to the theory are least controlled, so the quantitative theory-data agreement should be interpreted with caution.
Yet the range of theoretically possible dimensionalities at such $T$ spans only about $3$, only marginally different from the prediction for independent neurons~\cite{gao2017Theory}.
This narrow range means that current data cannot distinguish whether the measured dimensionality reflects the network structure beyond the mere existence of interactions.
Experiments for longer measurement times, more neurons, and with various perturbations are needed to place stronger constraints on network models.

There are ample experiments where dimensionality has been reported with the duration of behavior serving as the measurement time $T$~\cite{churchland2012Neural, mazzucato2016Stimuli, perich2018Neural, gallego2018Cortical, russo2020Neural, gallego2020Longterm, bartolo2020Dimensionality, snyder2021Stable, altan2023Lowdimensional, oby2025Dynamical}.
However, we see from the theory that the measurement time $T$ and the autocorrelation width $\tau_\acov$ jointly determine the dimensionality, while the autocorrelation width, or equivalently the form of autocorrelation, is usually not reported in experiments.
Direct comparison with the theory requires reporting both the measurement time and the autocorrelation width, together with the convention used to define $\tau_\acov$.
When an experiment uses a definition of $\tau_\acov$ different from \eqref{eq:tau-acov}, as in \citep{gao2017Theory}, the comparison is still possible provided that convention is reported explicitly, since the two definitions can then be made consistent by a simple rescaling.

Our analysis suggests reporting the dimensionality over multiple time windows, in addition to the longest $T$ as is currently common in the experimental literature.
This would provide a more complete description of network behavior across timescales, and comparison with theoretical predictions would identify the scale at which any disagreement arises.

\subsection*{Activity similarity between two behavioral contexts}\phantomsection\label{sec:two}

We now consider the similarity of the system's activity under two different, but possibly related, behavioral contexts.
To represent this setup in our model, we consider two copies of \eqref{eq:ode} that share a single coupling matrix $J$ but receive different external inputs $f_{1,i}$ and $f_{2,i}$.
The inputs remain unstructured across the population, while the two contexts' inputs may be correlated with each other.
To do this, we independently sample for each neuron the pair of external inputs as $\spmx{f_{1,i}\\f_{2,i}}\fl\oNor\ev{0,I^2\rho}$, where $\rho=\spmx{1&\rhoin\\ \rhoin&1}$ is the correlation matrix.
For simplicity, we take the two external inputs to have the same strength $I$, so that the corresponding activities $r_1$ and $r_2$ share the single-replica statistics of \nameref{sec:tld}.
Intuitively, if we approximate the network's activity under each external input as a shifted ellipsoid, cf.~Figure~\ref{fig:two}, then the two ellipsoids share the same size and roundness, representing the variance and dimensionality quantified by PR, and are shifted over the same distance from the state space's origin.
The two activities thus differ in the directions of their ordered responses and in the orientations of their temporal chaos.
In Figure~\ref{fig:two}, these correspond to the angle between the yellow crosses and the orientations of the blue ellipsoids.

\subsection*{Ordered responses preserve the similarity between behavioral contexts}\phantomsection\label{sec:two-bar}

\begin{figure}[bt!]
  \centering
  \includegraphics{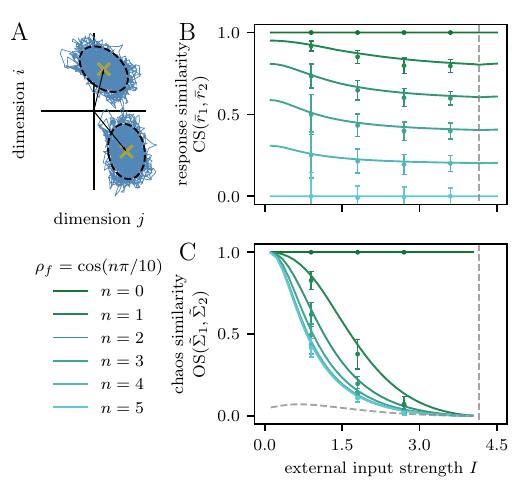}
  \caption{Similarity between activity under two behavioral contexts.
    A: Schematically, under each behavioral context, there is an ordered response around which temporal chaos fluctuates.
    B: The ordered-response similarity $\oCS$ remains close to the similarity between behavioral contexts $\rhoin$, decreases only weakly with external input strength, and quickly plateaus, with curves labeled by $\rhoin=\cos(n\pi/10)$.
    A small deviation visible near the chaos-to-order transition reflects the disappearance of temporal chaos.
    C: The orientation similarity $\oOS$ changes weakly at low external input strength, then decreases strongly with $I$ and approaches the random-orientation baseline $\oOS = \Tld{\oPR}_\infty$, marked by the dashed curve.
    All panels: The transition from chaotic to ordered dynamics is marked by the vertical dashed line.}
  \label{fig:two}
\end{figure}

The difference in direction between two arbitrary vectors $v_1$ and $v_2$ can be quantified by the cosine similarity $\oCS\ev{v_1,v_2}=v_1\cdot v_2/\lrpr{\lrvt{v_1} \lrvt{v_2}}$.
To the leading order in $\nni$, $\oCS\ev{f_1, f_2}$ is simply the correlation coefficient $\rhoin$:
\begin{equation}
  \label{eq:cs}
  \oCS\ev{\bar{r}_{1}, \bar{r}_{2}} = \frac{\bar{C}_{12}}{\bar{C}}, \quad\text{where}\quad \bar{C}_{12}=\frac{1}{\nni}\sm{\bar{r}_{1,i}\bar{r}_{2,i}}{i}
\end{equation}
is a new two-replica statistics extending the single-replica statistics $\bar{C}$ in \eqref{eq:op1-bar}, describing the overlap between the two replicas' ordered response.
The semi-analytic values of $\bar{C}_{12}$ come from the two-replica DMFT equation \eqref{eq:2-repl-dmft}, derived in \methref{sec:dmft2} via a saddle-point calculation.
Figure~\ref{fig:two}B shows the results.

At weak external input strengths $I\ll 1$, expanding the two-replica DMFT equation gives $\oCS=\rhoin$, so ordered responses preserve the similarity between the external inputs.
As $I$ increases, the parts of the inputs that differ between contexts drive neurons into different regions of the nonlinearity, reducing $\oCS$.
This decrease saturates once many neurons lie in low-gain regions.
Near the chaos-to-order transition, a weak local deviation appears as temporal chaos vanishes.
The saturation persists beyond the transition, as shown in \apperef{sec:cs-sat}, suggesting that neural populations can preserve input similarities with a bounded decrease.

The behavior of $\oCS$ is therefore mainly determined by where each neuron sits on the nonlinearity, a property of single neurons.
This is also consistent with the fact that the calculation needed for $\oCS$ does not exceed the saddle-point level in the replica calculation.
Comparing the measured $\oCS$ to our predictions therefore tests our single-neuron modeling assumptions.
Because finite measurement times affect ordered responses only weakly (see \eqref{eq:finite-cov}), our long-time prediction for $\oCS$ already applies at experimentally accessible $T$.
In experiments, quantities similar to $\oCS$ have been reported in the literature~\cite{tobin2024Distinct}, but for natural neural processes, external inputs to the system are often hidden, making similarities between them hard to estimate.
Modern stimulation techniques such as optogenetics allow controlling the similarity between inputs directly~\cite{liang2011Patterned, zhu2012Highresolution}, enabling this comparison.

\subsection*{Temporal-chaos orientation diverges between behavioral contexts}\phantomsection\label{sec:two-tld-long}

We now describe the relative orientation between the two temporal chaos $\tld{r}_1$ and $\tld{r}_2$.
Each is summarized by its covariance matrix, which can be visualized as an ellipsoid (Figure~\ref{fig:two}A).
Unlike a vector, an ellipsoid has multiple principal axes, and each axis is defined only up to sign.
This means we cannot use the simple cosine similarity to quantify the relative orientation between the two covariance matrices for temporal chaos $\tld{\Sigma}_{1}$ and $\tld{\Sigma}_{2}$, where we have suppressed the subscript $T\to\infty$.
One option for quantification is
\begin{equation}
  \label{eq:os}
  \oOS\ev{\Tld{\Sigma}_{1},\Tld{\Sigma}_{2}}
  = \frac{\oTr\ev{\Tld{\Sigma}_{1}\Tld{\Sigma}_{2}}}{ \sqrt{\oTr\ev{\tensor{\Tld{\Sigma}}{_{1}^2}}\oTr\ev{\tensor{\Tld{\Sigma}}{_{2}^2}}}},
\end{equation}
and we refer to it as the \emph{orientation similarity} (OS).
It generalizes the $\cos^2\theta$ similarity between two 1D orientations separated by angle $\theta$, and in higher dimensions it can be viewed as the intensive version of the \emph{shared dimensionality} between two ellipsoids~\cite{giaffar2023Effective}.
$\oOS$ takes value between $0$ and $1$ as for $\cos^2\theta$, and as the shared dimensionality, $\oOS=0$ represents no shared dimensions, and $\oOS=1$ represents full overlap.
Two ellipsoids can share orientation by chance, so the natural baseline for $\oOS$ is its expected value under random orientations: $\ep{\oOS\ev{\Tld{\Sigma}_{1},O\Tld{\Sigma}_{2} O^T}}{O} = \sqrt{\Tld{\oPR}_1\Tld{\oPR}_2}$, with $O$ drawn from the Haar measure on orthogonal matrices.
When both contexts have the same external input strength $I$, the two single-replica PRs match, and this baseline simplifies to $\Tld{\oPR}_\infty$ (see \nameref{sec:tld-long}).

For $\oOS$ in \eqref{eq:os}, the denominator is the single-replica quantity in \eqref{eq:cov-tld-sq}, while the numerator is related to fluctuations of the two-replica autocovariance
\begin{equation}
  \label{eq:op2-tld}
  \tld{C}_{12t_1t_2}=\frac{1}{\nni}\sm{\tld{r}_{1it_1}\tld{r}_{2it_2}}{i}
\end{equation}
around its saddle-point value $0$, because the temporal-chaos dynamics of the two replicas decouple.
The final expression for $\oOS$ is
\begin{equation}
  \label{eq:os-tld-result}
  \oOS =\frac{\ep{\tld{C}_{12t,t+\infty}^2}{t}}{\ep{\tld{C}_{t,t+\infty}^2}{t}}.
\end{equation}

The variance of $\tld{C}_{12}$, given by \eqref{eq:2-repl-tld-inv-non-non} from \methref{sec:dmft2}, yields the results in Figure~\ref{fig:two}C.
When the two external inputs are identical, $\oOS=1$.
As they become large and distinct, $\oOS$ decreases toward the random-orientation baseline $\Tld{\oPR}_\infty$ (non-monotonic dashed curve), with most of the decrease at moderate input strengths where $\Tld{\oPR}_\infty$ also declines.
At weak inputs, chaotic fluctuations remain aligned across contexts in the fluctuation-dissipation regime; at stronger inputs, the two contexts suppress different subsets of neurons and recurrent interactions reshape the remaining fluctuations along different directions.
$\oOS$ thus behaves differently from the quickly-saturating $\oCS$: even at $\rhoin \approx 0.95$ (ordered responses highly similar), temporal-chaos orientations can decorrelate substantially at larger $I$.
Past the chaos-to-order transition (vertical dashed line), temporal chaos vanishes and $\oOS$ is no longer defined.

Since experiments have a finite measurement time $T$, we define $\oOS_T$ by evaluating \eqref{eq:os} at the two finite-time covariances $\tld{\Sigma}_{1 T t_\mea }$ and $\tld{\Sigma}_{2 T t_\mea }$.
Using the same approximations as in \eqref{eq:finite-pr-tld-result}, and noting that the two replicas have independent temporal-chaos fluctuations, we show in \methref{sec:finite} that
\begin{equation}
  \label{eq:finite-os-result}
  \frac{\oOS_{T}}{\oOS_{\infty}} = \frac{\Tld{\oPR}_{T}}{\Tld{\oPR}_{\infty}}.
\end{equation}
Compared to the random baseline $\Tld{\oPR}_T$, $\oOS_{T}$ therefore differs by the fixed long-time factor $\oOS_{\infty}/\Tld{\oPR}_\infty$.

Unlike $\oCS$, $\oOS$ depends on inter-neuron correlations in addition to single-neuron properties, so comparing measured $\oOS$ to our prediction can reveal structures in real neural systems not captured by the minimal network.
In particular, it could be beneficial to observe whether the decrease in $\oOS_T$ over external input strength $I$ coincides with the decrease in $\Tld{\oPR}_T$.
A disagreement with this prediction would indicate that real interactions have structure making correlations robust to gain changes, beyond what the minimal network captures.
As for $\Tld{\oPR}_T$, we expect $\oOS_T$ to be sensitive to single-neuron details of the model when the measurement time is relatively short $T\lesssim\tau_\acov$, and as the measurement time gets longer $T\gtrsim\tau_\acov$, the details will be dominated by correlations, reflecting actual computation.

\subsection*{Geometry over multiple behavioral contexts}\phantomsection\label{sec:bar}

We now consider the geometry of neural activity over multiple behavioral contexts, modeled by driving the same network (fixed coupling matrix $J$) with distinct external inputs.
Each input results in a different ordered response, and intuitively, this leads to a cloud of centers for temporal chaos, which we can approximate as an ellipsoid through its covariance matrix, illustrated in Figure~\ref{fig:bar}A.
Its size is given by $\bar{C}$, and we are now interested in its roundness, representing the dimensionality of activity over the sampled behavioral contexts.
We refer to this dimensionality as the \emph{multi-context dimensionality}.
Further, since the temporal chaos around each ordered response has a different orientation, in this section we ignore the complex orienting behavior (shown by the faint colors in Figure~\ref{fig:bar}A).

We again quantify the multi-context dimensionality using the PR evaluated at the covariance $\bar{\Sigma}$ of the ordered response over the collection of external inputs,
\begin{equation}
  \label{eq:pr-bar}
  \Bar{\oPR}\ev{\bar{\Sigma}} = \frac{\lrpr{\oTr\ev{\bar{\Sigma}}}^2}{\nni\oTr\ev{\bar{\Sigma}\idc{^2}}}, \quad\text{where}\quad \bar{\Sigma}_{ij} = \ep{\bar{r}_{i}\bar{r}_{j}}{f}.
\end{equation}
Similar to \eqref{eq:pr-tld}, the trace in the numerator is $\nni\bar{C}$ as in \eqref{eq:op1-bar}, and the trace in the denominator is given by 
\begin{equation}
  \label{eq:cov-bar-sq}
  \begin{aligned}
    \oTr\ev{\bar{\Sigma}\idc{^2}}
    &= \sm{\ep{\bar{r}_{1,i}\bar{r}_{1,j}\bar{r}_{2,i}\bar{r}_{2,j}}{f_{1}f_{2}}}{ij}
      = \nni^2\ep{\bar{C}_{12}^2}{f_{1}f_{2}}.
  \end{aligned}
\end{equation}

Equation~\ref{eq:cov-bar-sq} only requires the external inputs $\lrpr{f_{1,i},\cdots,f_{\nnc,i}}$ to be i.i.d.\ over neuron index $i$, leaving the joint distribution across contexts unconstrained.
For simplicity, we take the contexts to be independent.
We then evaluate \eqref{eq:cov-bar-sq} either as a true expectation in the large-context-number limit or as an empirical average for finitely many contexts.

\subsection*{The large-context-number dimensionality increases with external input strength}\phantomsection\label{sec:bar-many}

For independent external inputs, the pair similarity is $\rhoin=0$ when $f_1\neq f_2$.
In the large-context-number limit $\nnc\to\infty$, by an approximation analogous to \eqref{eq:cov-tld-sq}, the expectation $\ep{\bar{C}_{12}^2}{f_{1}f_{2}}$ in \eqref{eq:cov-bar-sq} reduces to the $\sim 1/\nni$ saddle-point variance of $\bar{C}_{12}$ around zero, over uncorrelated input pairs.
We access this variance using a two-replica calculation around the saddle-point, detailed in \methref{sec:dmft2}, and the resulting expression for the multi-context dimensionality over large context numbers $\nnc\to\infty$ is
\begin{equation}
  \label{eq:pr-bar-result}
  \Bar{\oPR}_\infty = \frac{\bar{C}^2}{\nni\ep{\bar{C}_{12}^2}{f_{1}f_{2}}} = \lrpr{1-g^2\ep{\phi'\ev{h_{it}}}{it}^2}^2,
\end{equation}
where $\ep{\phi'\ev{h_{it}}}{it}$ is the gain averaged over both population and time.

\begin{figure}[bt!]
  \centering
  \includegraphics{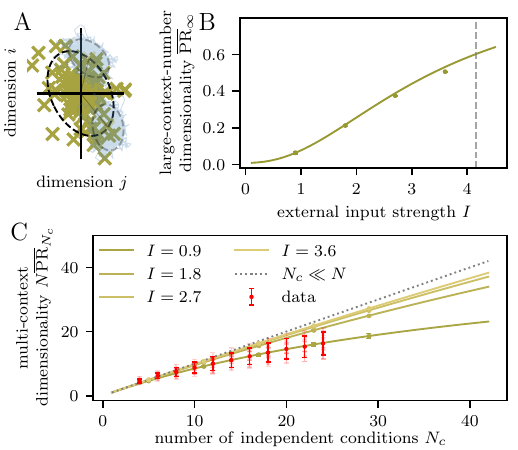}
  \caption{Statistics of activity over multiple behavioral contexts.
    A: Schematic.
Each behavioral context produces an ordered response; the cloud of ordered responses across contexts is approximated as an ellipsoid.
    B: $\Bar{\oPR}_\infty$ increases monotonically with external input strength $I$, with slow growth at small $I$ and a decreasing growth rate at large $I$.
    The vertical dashed line marks the transition from chaotic to ordered dynamics.
    C: The dependence of the multi-context dimensionality $\Bar{\oPR}_\nnc$ on context number $\nnc$ mirrors that of $\Tld{\oPR}_T$ on measurement time $T$.
    The large-neuron-number limit $\nnc\ll\nni$ is shown in dashed lines, where the dimensionality increases linearly.
    Otherwise the dimensionality grows sublinearly with rates dependent on the external input strength $I$.
    Digitized dimensionality measurements from \citep{bartolo2020Dimensionality} are shown in red after converting task block number to context number by assigning two behavioral contexts to each block; see \methref{sec:numerics}.
  }
  \label{fig:bar}
\end{figure}

$\Bar{\oPR}_\infty$ grows monotonically with external input strength $I$, as shown by the semi-analytic curves in Figure~\ref{fig:bar}B.
At weak input strength $I\ll 1$, growth is slow.
In this fluctuation-dissipation-type regime, the ordered responses are biased toward directions where autonomous temporal chaos has large fluctuations, so the response cloud occupies only a low-dimensional set of directions, consistent with \eqref{eq:pr-bar-result}.
As $I$ increases, saturation caps the amplified recurrent response, and the high-dimensional raw input contributes more strongly to the ordered responses, increasing $\Bar{\oPR}_\infty$.
The prediction agrees with simulations except near $I=4$, where higher-dimensional spaces become harder to sample numerically.

Although $\Bar{\oPR}_\infty$ might seem harder to measure experimentally than $\Tld{\oPR}_\infty$, accurate measurement of ordered responses requires only recording times of order $\tau_\acov$, far shorter than what temporal chaos requires.
The difficulty lies in preparing a large number of independent behavioral contexts with similar levels of input from external regions.
This could be hard for natural contexts, but feasible for artificial contexts, for example by using opto-genetics stimulation \cite{liang2011Patterned, zhu2012Highresolution}.

\subsection*{Finite context number is analogous to finite measurement time}\phantomsection\label{sec:bar-few}

The finite-context-number dimensionality $\Bar{\oPR}_{\nnc}$ is of interest both because the large-context-number limit is hard to measure experimentally, and because we may want to characterize how a system represents a finite set of sampled behavioral contexts.
We show in \methref{sec:finite}, that
\begin{equation}
  \label{eq:finite-pr-bar-result}
  \Bar{\oPR}_{\nnc}= \frac{\bar{C}^2}{\nni\ep{\bar{C}_{12}^2}{f_{1}f_{2}}+\frac{\nni}{\nnc}\bar{C}^2}.
\end{equation}
This has the same form as the finite-time dimensionality in \eqref{eq:finite-pr-tld-result-approx}, when the measurement window spans several autocorrelation times.
The mapping is $\tld{C}_0\mapsto\bar{C}$, $\ep{\tld{C}_{t,t+\infty}^2}{t}\mapsto\ep{\bar{C}_{12}^2}{f_{1}f_{2}}$, and $T/\tau_\acov\mapsto\nnc$.
Structurally, the two situations differ: finite context number involves independent samples, while finite measurement time involves temporally correlated samples.
Algebraically they are analogous, both being finite-sampling effects where a subset of states occupies fewer dimensions than the full neural manifold, yielding the same functional form for the participation-ratio correction.

The results are shown in Fig.~\ref{fig:bar}C, and similar to the relationship between \eqref{eq:pr-tld-result} and \eqref{eq:finite-pr-tld-result}, \eqref{eq:finite-pr-bar-result} adds a correction to the large context-number dimensionality in \eqref{eq:pr-bar-result}.
Analogous to $\Tld{\oPR}_T$, for $\nnc\gg\nni$, the dimensionality $\Bar{\oPR}_\nnc$ approaches its limit $\Bar{\oPR}$ from below as $\nni/\nnc$, and for $\nnc\lesssim\nni$, the growth of $\Bar{\oPR}_\nnc$ is linear if $1<\nnc\ll\nni$ is allowed by the size of $\nni$ and sublinear otherwise.
Unlike the temporal-chaos dimensionality, this multi-context dimensionality is controlled by ordered-response overlaps, which are much more self-averaging and therefore less affected by finite-size concerns near the transition.

We compare this prediction with the linear dimensionality measurements of \citep{bartolo2020Dimensionality}, defined as the number of principal components needed to explain $80\%$ of the variance in trial-averaged responses, across increasing numbers of task blocks.
Because each block in the task considered in \citep{bartolo2020Dimensionality} contains two unique images, we convert block number to context number by assigning two behavioral contexts to each block.
We pool the four reported conditions because the comparison concerns the overall range of experimental dimensionality values across the task.
The resulting processed data are shown with the theory curves in Figure~\ref{fig:bar}C.
Under this block-to-context mapping, the experimental dimensionalities are close to the minimally structured prediction.
Assuming this interpretation of behavioral context, the agreement suggests that additional structure is not required to explain the measured multi-context dimensionality.
This comparison demonstrates that $\Bar{\oPR}_\nnc$ can be used to relate finite-context experimental dimensionality measurements to a minimally structured network baseline.

\section*{Discussion}

Here we explored whether low-dimensional population activity implies special structure in the underlying circuit.
To do this, we evaluated the dimensionality of activity in a minimally structured random recurrent network and asked if this dimensionality is already as low as experimental values.
Prior DMFT theory of such networks worked in the infinite-measurement-time limit without external inputs; for a quantitative comparison with experiments we additionally incorporated finite measurement time, external inputs, and finite system size, all matching the constraints of real recordings.
The model then accounts quantitatively for the low dimensionality reported in data.
Over the experimental range of $T/\tau_\acov$, however, the predicted dimensionality varies only weakly, so current measurements cannot distinguish generic random interactions from additional circuit or task structure.
The model also disagrees with data at shorter timescales, but this likely reflects single-neuron physiological details irrelevant to network-wide computation.
Our results therefore suggest that low dimensionality for finite measurement times alone is insufficient to establish whether the underlying circuit has additional structure beyond random interactions.

To distinguish data that genuinely requires structure beyond random interactions, we identified additional geometric quantities predicted by the same minimally structured network.
External input changes both the variance and the geometry of temporal chaos.
The long-time dimensionality of temporal chaos varies non-monotonically with input strength, rising at weak input and falling at strong input.
Across behavioral contexts, the ordered responses stay close to the input direction, while the geometry of the temporal fluctuations changes much more strongly: orientation similarity falls rapidly toward the random-overlap baseline as input strength grows.
Multi-context dimensionality, finally, varies systematically with both input strength and the number of contexts.
Together, these predictions specify which features of neural geometry would be unsurprising in a minimally structured network, and which would indicate additional structure.

To achieve these results, we made several assumptions.
First, we assumed the network is self-averaging, as is standard in DMFT, so that population statistics are dominated by their typical values, with realization-to-realization fluctuations subleading.
This assumption breaks down as the network approaches the chaos-to-order transition; fitting the random model in practice should therefore be followed by checking whether the fitted model at the assumed $\nni$ is as self-averaging as intended.
Second, we assumed time-independent external inputs within a given behavioral context.
For a linear network, arbitrary time-dependent inputs could be handled by decomposing the input into temporal modes and superposing the corresponding responses, but for nonlinear networks no such decomposition exists, so a particular time-dependent input is informative only when its temporal structure is well constrained by the experiment.
Third, we did not treat subsampling explicitly: in general, the dimensionality measured from a recorded subset can differ from that of the full population, depending on both how many neurons are sampled and how.
At short measurement times, however, we showed that the activity occupies a low-dimensional linear subspace, so approximating subsampling by random projections does not strongly distort the measured geometry~\cite{gao2017Theory}.

We have chosen the participation ratio to quantify dimensionality, and that choice has two consequences.
First, PR counts all dimensions and weights them by variance, so directions with very small variance contribute little.
Its relation to PCA thresholding depends on the eigenspectrum: a high explained-variance threshold can count many low-variance directions and give a larger dimension than PR, whereas a low threshold can miss broader covariance structure and give a smaller one.
When the spectrum has a natural cutoff, the two measures converge.
Second, PR is a linear measure: when the activity manifold is strongly curved in state space, a linear quantification can exceed the intrinsic dimensionality~\cite{altan2021Estimating, gulati2026latent}.
Manifolds relevant for neural computation are thought to often be smooth in state space~\cite{chung2021Neural}, in which case any overestimation should be moderate.
We use PR because it is analytically tractable and stays close to PCA-based quantities commonly reported in experiments.

Overall, our work shows that low dimensionality should not by itself be taken as evidence for specially organized circuit structure.
To establish whether the network is more than random, experiments need either to probe regimes where the random predictions vary appreciably (for example at longer measurement times) or to measure additional geometric quantities beyond dimensionality.
We propose a specific set of such quantities: the non-monotonic dependence of temporal-chaos dimensionality on input strength, the rapid decay of orientation similarity across contexts, and the multi-context dimensionality as a function of context number.
Each is quantitatively predicted by a minimally structured network, so any deviation would indicate additional structure.
If future measurements show behavior incompatible with the minimally structured baseline, additional structure is needed, and one can either introduce it ad hoc or compare to the richer structured models in the literature~\cite{mastrogiuseppe2018Linking, aljadeff2015Transition, marti2018Correlations, clark2024Connectivity}.

\section*{Acknowledgments}
We are grateful to David Clark and Audrey Sederberg for stimulating discussions.
This work was supported, in part, by the Simons Foundation Investigator award to I.\ N.\ and by the NIH grants R01-NS084844 and R01-NS099375.

\balance
\bibliographystyle{unsrtnat}
\bibliography{non-EI.bib}

\onecolumn
\section*{Methods}

In this section we show the details of how we calculate or compute the quantities presented in Results.
Sections \nameref{sec:dmft1-sp}, \nameref{sec:dmft1-fluct}, and \nameref{sec:dmft2} cover the replica method from which the long-time quantities are calculated.
More specifically, the variance of the ordered response and temporal chaos is obtained from \nameref{sec:dmft1-sp}, the dimensionality of temporal chaos from \nameref{sec:dmft1-fluct}, and the dimensionality of ordered responses and the similarity values from \nameref{sec:dmft2}.
\nameref{sec:finite} covers the method to calculate quantities over finite time or behavioral contexts.
And \nameref{sec:numerics} covers numeric details of the simulation, semi-analytic calculation, and extraction of experimental data.

\subsection*{DMFT for variance and autocovariance}\phantomsection\label{sec:dmft1-sp}
The derivation in this section loosely follows \citep{crisanti2018Path}, with the main difference that we introduce two order parameters, $\bar{C}$ and $\tld{C}_\tau$, rather than one.
We include the details here for completeness.
In the replica method, one hopes to obtain values of various statistics by working with the free energy.
We specifically focus on the variance of the ordered response $\bar{C}$ and the autocovariance of temporal chaos $\tld{C}_\tau$.
Since the free energy cannot be evaluated directly because of the nonlinearity $\phi$, we recast it as a saddle-point integral over the auxiliary order parameters $\bar{C}$ and $\tld{C}_\tau$, which become tractable in the large-$\nni$ limit.

\subsubsection*{Decoupling dynamics over space}

We start by writing down the path integral over all trajectories of the preactivation over time, given a particular quenched disorder $q$, which includes the connectivity $J_{ij}$, external input $f_i$, and initial condition $\sta{h}_{i,t=0}$:
\begin{align}
  \label{eq:1-repl-zq}
  \begin{aligned}
    Z_q
    &= \int \pd{\dd h_{it}}{it} \pd{\ddelta\ev{h_{it}-\sta{h}_{q,i,t}}}{it} \ee^{\ii\sm{\nt{\crc{b}_{it}h_{it}}{t}}{i}}
    \\ &=\int \pd{\dd h_{it}}{it}\Pd{\ddelta\Ev{\dd t \Lrpr{\lrpr{h_{it}+\ddel_th_{it}} -\sm{J_{ij}\phi\ev{h_{j,t-\dd t}}}{j} -f_i -b_{it}}}}{it} \ee^{\ii\sm{\nt{\crc{b}_{it}h_{it}}{t}}{i}}
    \\ &=\int \Pd{\frac{\dd h_{it}\dd \crc{h}_{it}}{2\pi}}{it} \exp\Ev{\ii\Sm{\Nt{\mtx{-\crc{h}_{it}\lrpr{h_{it}+\ddel_th_{it}}\\ +\crc{h}_{it}b_{it}+\crc{b}_{it}h_{it}}}{t}}{i}} \\&\qquad\times\exp\Ev{\ii\sm{\nt{\crc{h}_{it}\phi\ev{h_{jt}}}{t}J_{ij}}{ij}}\exp\Ev{\ii\sm{\nt{\crc{h}_{it}}{t}f_i}{i}}.
  \end{aligned}
\end{align}
The bracketed expression in the last equality is split across two lines for layout only. The line break does not indicate a vector or matrix.
The neural index $i=1,\cdots,\nni$, and the time $t$ is discretized into $t=\dd t,\cdots,\nnt\dd t=T_\infty$, where $\dd t$ is the step size.
For the later analysis, we include possible perturbations $b_{it}$s on the r.h.s.\ of the differential equation \eqref{eq:ode} at time $t$ to neuron $i$, and $\sta{h}_{q,i,t}$ then represents the true solution to the perturbed differential equation.
By convention, we also include $\crc{b}_{it}$ as the current for the preactivations, but we could instead have currents for $\bar{C}$ and $\tld{C}_\tau$.
The second line of \eqref{eq:1-repl-zq} replaces $\ddelta\ev{h - \sta{h}}$ with the equation-of-motion form of the Dirac delta. 
The composition rule $\ddelta(f(x)) = \ddelta(x)/|f'(x)|$ produces the factors of $\dd t$ (Jacobian of the time discretization, interpreted in Ito's scheme).
We then express the $\ddelta$ functions in Fourier space in the third line, since we will later average away the quenched disorder $q$ with a Gaussian integral.

To turn the partition function into a free energy, we assume the system is replica symmetric, i.e., for replica number $n$
\begin{equation}
  \label{eq:rep-sym}
  \begin{gathered}
    \ep{Z_q^n}{q} = \ep{Z_{q}}{q}^n,\\
    F = \log\ep{Z_q}{q}.
  \end{gathered}
\end{equation}
We proceed under the replica symmetry assumption, standard in the chaotic phase, where the saddle-point solution is symmetric under permutation of replica indices; this can break near the transition to ordered dynamics.
Now the average directly acts on $Z_q$ in \eqref{eq:1-repl-zq}, and we can perform the Gaussian averages over the connectivity $J$ and external input $f$ since the action is linear in them.
The initial condition $\sta{h}_{it_0}$ is on the other hand hard to average away due to the presence of $\phi$, but it is not essential since we believe the system is ergodic and we are not interested in initial transients.
After the average,
\begin{equation}
  \label{eq:1-repl-f}
  \begin{aligned}
    F=& \log \int \Pd{\frac{\dd h_{it}\dd \crc{h}_{it}}{2\pi}}{it}\prod_{i}\Lbk\exp\Ev{\ii\Nt{\mtx{-\crc{h}_{it}\lrpr{h_{it}+\ddel_th_{it}}\\+\crc{h}_{it}b_{it}+\crc{b}_{it}h_{it}}}{t}}\\&\qquad\times\exp\Ev{-\frac{1}{2}\Nt{\crc{h}_{it_1}\crc{h}_{it_2}\Lrpr{g^2\Sm{\frac{1}{\nni}\phi\ev{h_{jt_1}}\phi\ev{h_{jt_2}}}{j}+I^2}}{t_1t_2}}\Rbk,
  \end{aligned}
\end{equation}
and we can see that, conveniently, different neural indices are only coupled through $\sm{\phi\ev{h_{jt_1}}\phi\ev{h_{jt_2}}}{j}/\nni = \bar{C} +\tld{C}_{t_1t_2}$.
This means that, by introducing $\bar{C} = \sm{\ep{\phi\ev{h_{it}}}{t}^2}{i}/\nni$ and $\tld{C}_{t_1t_2} = \sm{\phi\ev{h_{it_1}}\phi\ev{h_{it_2}}-\ep{\phi\ev{h_{it}}}{t}^2}{i}/\nni$ as variables, we can simultaneously obtain their values and decouple the dynamics spatially
\begin{equation}
  \label{eq:1-repl-od-trick}
  \begin{aligned}
    F =& \log \int \Pd{\frac{\dd h_{it}\dd \crc{h}_{it}}{2\pi}}{it} \int\dd\bar{C}\, \ddelta\Ev{\bar{C} - \frac{1}{\nni} \sm{\ep{\phi\ev{h_{it}}}{t}^2}{i}} \\&\qquad \int\pd{\dd \tld{C}_{t_1t_2}}{ \ss{t_1t_2\\t_1\leq t_2}} \Pd{\ddelta\Ev{\tld{C}_{t_1t_2} -\frac{1}{\nni} \Sm{\phi\ev{h_{it_1}}\phi\ev{h_{it_2}} - \ep{\phi\ev{h_{it}}}{t}^2}{i}}}{ \ss{t_1t_2\\t_1\leq t_2}}    \\&\qquad\times\prod_{i}\Lbk\exp\Ev{\ii\Nt{\mtx{-\crc{h}_{it}\lrpr{h_{it}+\ddel_th_{it}}\\+\crc{h}_{it}b_{it}+\crc{b}_{it}h_{it}}}{t}}    \\&\qquad\qquad\times\exp\Ev{-\frac{1}{2}\Nt{\crc{h}_{it_1}\crc{h}_{it_2}\Lrpr{g^2\lrpr{\bar{C} +\tld{C}_{t_1t_2}}+I^2}}{t_1t_2}}\Rbk.
  \end{aligned}
\end{equation}
Note that we only introduced $\nnt^2/2$ delta functions for $\tld{C}$, since $\tld{C}_{t_1t_2}=\tld{C}_{t_2t_1}$ by definition and one of them is a redundant degree of freedom.
Alternatively, one could introduce $\tld{C}_{t_1t_2}$ and $\tld{C}_{t_2t_1}$ separately with $\nnt^2$ delta functions, and then, in later subsections, when calculating fluctuations around the saddle point, force each derivative with respect to one to also act on the other, and take the long-lag limit $t_2-t_1\to\infty$ before inverting the Hessian for the fluctuation~\cite{clark2024Connectivity}.

We again express the Dirac deltas $\ddelta$ in Fourier space, since this allows us to reorder the integrals in a form where different integrals over $h_{it}$ and $\crc{h}_{it}$ are independent for different $i$-s, and all of them depend on the value of the statistics $\bar{C}$ and $\tld{C}_{t_1t_2}$.
Additionally, we constrain ourselves to only consider spatially uniform perturbations and currents ($b_{it}=b_{t}$ and $\crc{b}_{it}=\crc{b}_{t}$ for all $i$) since this is all we need in later calculations, and this simplifies the decoupled integrals to be identical.
\begin{equation}
  \label{eq:1-repl-od}
  \begin{aligned}
    F=& \log \int\frac{\dd\bar{C}\dd\crc{\bar{C}}}{2\pi/\nni}\exp\ev{-\ii \nni\crc{\bar{C}}\bar{C}}\int\Pd{\frac{\dd \tld{C}_{t_1t_2}\dd \crc{\tld{C}}_{t_1t_2}}{2\pi/\nni}}{ \ss{t_1t_2\\t_1\leq t_2}}\exp\ev{-\ii \nni\sm{\crc{\tld{C}}_{t_1t_2}\tld{C}_{t_1t_2}}{ \ss{t_1t_2\\t_1\leq t_2}}}    \\&\qquad\Lbk\int \Pd{\frac{\dd h_{t}\dd \crc{h}_{t}}{2\pi}}{t} \exp\Ev{\ii\Sm{\crc{\tld{C}}_{t_1t_2}\lrpr{\phi\ev{h_{t_1}}\phi\ev{h_{t_2}} - \ep{\phi\ev{h_{t}}}{t}^2}}{ \ss{t_1t_2\\t_1\leq t_2}}+\ii\crc{\bar{C}}\ep{\phi\ev{h_{t}}}{t}^2}    \\&\qquad\qquad\times\exp\Ev{\ii\Nt{\mtx{-\crc{h}_{t}\lrpr{h_{t}+\ddel_th_{t}}\\+\crc{h}_{t}b_{t}+\crc{b}_{t}h_{t}}}{t}}    \\&\qquad\qquad\times\exp\Ev{-\frac{1}{2}\Nt{\crc{h}_{t_1}\crc{h}_{t_2}\Lrpr{g^2\lrpr{\bar{C} +\tld{C}_{t_1t_2}}+I^2}}{t_1t_2}}\Rbk^{\nni}.
  \end{aligned}
\end{equation}

\subsubsection*{DMFT equations}

We have now successfully simplified the spatial dimension of the dynamics from $\nni$ to $1$ (or $4$ including the auxiliary conjugate fields), and we now try to solve for the statistics $\bar{C}$ and $\tld{C}_\tau$ in this simpler dynamics.
As expected, the integrals over $h_{t}$ and $\crc{h}_{t}$ do not appear easy to evaluate due to the nonlinearity $\phi$, but fortunately, the integrals over $\bar{C}$ and $\tld{C}_\tau$ can be simplified with saddle-point approximations in the $\nni\to\infty$ limit.
We first rewrite the free energy as a hierarchy of free energies:
\begin{equation}
  \label{eq:1-repl-hierarchy}
  \begin{aligned}
    F_{\bar{\nam{C}}} =& \log \int\frac{\dd\bar{C}\dd\crc{\bar{C}}}{2\pi/\nni} \exp\ev{-\ii \nni\crc{\bar{C}}\bar{C} +F_{\tld{\nam{C}}}}
    \\F_{\tld{\nam{C}}} =&\log\int\Pd{\frac{\dd \tld{C}_{t_1t_2}\dd \crc{\tld{C}}_{t_1t_2}}{2\pi/\nni}}{ \ss{t_1t_2\\t_1\leq t_2}}\exp\ev{-\ii \nni\sm{\crc{\tld{C}}_{t_1t_2}\tld{C}_{t_1t_2}}{ \ss{t_1t_2\\t_1\leq t_2}}+\nni F_{\nam{h}}}
    \\F_{\nam{h}} =& \log\int \Pd{\frac{\dd h_{t}\dd \crc{h}_{t}}{2\pi}}{t} \exp\Ev{\ii\Sm{\crc{\tld{C}}_{t_1t_2}\lrpr{\phi\ev{h_{t_1}}\phi\ev{h_{t_2}} - \ep{\phi\ev{h_{t}}}{t}^2}}{ \ss{t_1t_2\\t_1\leq t_2}}+\ii\crc{\bar{C}}\ep{\phi\ev{h_{t}}}{t}^2} \\&\qquad\qquad\times\exp\Ev{\ii\Nt{\mtx{-\crc{h}_{t}\lrpr{h_{t}+\ddel_th_{t}}\\+\crc{h}_{t}b_{t}+\crc{b}_{t}h_{t}}}{t}}    \\&\qquad\qquad\times\exp\Ev{-\frac{1}{2}\Nt{\crc{h}_{t_1}\crc{h}_{t_2}\Lrpr{g^2\lrpr{\bar{C} +\tld{C}_{t_1t_2}}+I^2}}{t_1t_2}},
  \end{aligned}
\end{equation}
where the original full free energy equals $F_{\bar{\nam{C}}}$ on the highest level.
In the $\nni\to\infty$ limit, since the action in $F_{\tld{\nam{C}}}$ is $\sim \nni$, $\tld{C}_{t_1t_2}$ has $\sim \pm 1/\sqrt{\nni}$ fluctuations around its $\sim_\nni 1$ saddle-point value, and $F_{\tld{\nam{C}}}$ is also $\sim \nni$.
Consequently, the action in $F_{\bar{\nam{C}}}$ is also $\sim \nni$, leading to $\sim \pm 1/\sqrt{\nni}$ fluctuations around its $\sim_\nni 1$ saddle-point value.

By the property of free energies, the saddle-point values of $\bar{C}$ and $\crc{\bar{C}}$ (abbreviated as $\dbl{\bar{C}}$, similarly for $\tld{C}$ and $h$) are given by the DMFT equations
\begin{equation}
  \label{eq:1-repl-dmft-bar}
  \begin{gathered}
    \ii\crc{\bar{C}}= -\frac{1}{2}g^2 \nt{\ep{\ep{\crc{h}_{t_1}\crc{h}_{t_2}}{\dbl{h}}}{\dbl{\tld{C}}}}{t_1t_2} \approx -\frac{1}{2}g^2 \nt{\ep{\crc{h}_{t_1}\crc{h}_{t_2}}{\dbl{h}}}{t_1t_2},
    \\ \bar{C} =  \ep{\ep{\ep{\phi\ev{h_{t}}}{t}^2}{\dbl{h}}}{\dbl{\tld{C}}}\approx\ep{\ep{\phi\ev{h_{t}}}{t}^2}{\dbl{h}},
  \end{gathered}
\end{equation}
where the approximation uses the fact that $\dbl{\tld{C}}_{t_1t_2}$ is tightly distributed.
Under this value of $\dbl{\bar{C}}$, the saddle-point value of $\dbl{\tld{C}}_{t_1t_2}$ is given by the DMFT equations
\begin{equation}
  \label{eq:1-repl-dmft-tld}
  \begin{gathered}
    \ii\crc{\tld{C}}_{t_1t_2}=-\dd t^2 g^2\ep{\crc{h}_{t_1}\crc{h}_{t_2}}{\dbl{h}},
    \\ \tld{C}_{t_1t_2} = \ep{\phi\ev{h_{t_1}}\phi\ev{h_{t_2}}}{\dbl{h}} - \ep{\ep{\phi\ev{h_{t}}}{t}^2}{\dbl{h}},
  \end{gathered}
\end{equation}
where note that each $\tld{C}_{t_1t_2}$ also appears in its other form $\tld{C}_{t_2t_1}$.
To evaluate the expectations over $\dbl{h}$, we rewrite the free energy for $\dbl{h}$ as
\begin{equation}
  \label{eq:1-repl-equiv-dyn}
  \begin{aligned}
    F_{\nam{h}} =&\log\int \Pd{\frac{\dd h_{t}\dd \crc{h}_{t}}{2\pi}}{t}  \exp\Ev{\ii\Sm{\crc{\tld{C}}_{t_1t_2}\lrpr{\phi\ev{h_{t_1}}\phi\ev{h_{t_2}} - \ep{\phi\ev{h_{t}}}{t}^2}}{ \ss{t_1t_2\\t_1\leq t_2}}+\ii\crc{\bar{C}}\ep{\phi\ev{h_{t}}}{t}^2} \\&\qquad\times\Ep{\exp\Ev{\ii\Nt{\mtx{-\crc{h}_{t}\lrpr{h_{t}+\ddel_th_{t}}\\+\crc{h}_{t}b_{t} +\crc{b}_{t}h_{t}}}{t} +\ii\Nt{\crc{h}_{t}\lrpr{\bar{\xi} +\tld{\xi}_{t}}}{t}}}{\ss{\bar{\xi}\fl \oNor\ev{0,g^2\bar{C}+I^2}\\\tld{\xi}_{t}\fl \oNor\ev{0,g^2\tld{C}_{t_1t_2}}}},
  \end{aligned}
\end{equation}
where the interaction terms are rewritten as an average over Gaussian variables $\bar{\xi}_i$ and $\tld{\xi}_{it}$ whose cumulants depend on $\bar{C}$ and $\tld{C}_{t_1t_2}$.
Assuming $\crc{\bar{C}}=\crc{\tld{C}}_{t_1t_2}=0$, \eqref{eq:1-repl-equiv-dyn} can be interpreted as the free energy of a 1D Langevin dynamics where the noise ($\tld{\xi}_t$ thermal and $\bar{\xi}$ quenched) is described by the statistic variables $\bar{C}$ and $\tld{C}_{t_1t_2}$.
And this assumption can be shown to be self-consistent, by showing the DMFT equations would indeed return $\crc{\bar{C}} = \crc{\tld{C}}_{t_1t_2} = 0$, using the trick $\ep{\crc{h}_{t_1}\crc{h}_{t_2}}{\dbl{h}} =\ddel_{b_{t_1}}\ddel_{b_{t_2}}\ep{1}{\dbl{h}}/\lrpr{\ii\dd t}^2$ which turns $\crc{h}_t$ into a Dirac delta (localized and normalized) perturbation at time $t$ to the Langevin dynamics.
Given this interpretation of $F_{\nam{h}}$, we write $\ep{f}{\xi} = \ep{\ep{f}{\tld{\xi}}}{\bar{\xi}}$ instead of $\ep{f}{\dbl{h}}$ in the following.

\subsubsection*{Solving DMFT equations}

In this subsubsection, we include additional details for solving the DMFT equations, using the re-interpretation of the saddle-point-approximated free energy.

Since the statistics $\ep{h_th_{t+\tau}}{t}$ of the equivalent system should also be stationary, we write this 1D Langevin dynamics in Fourier space and use the convolution theorem to get the dynamics $(1-\ddel_{\tau}^2)\ep{h_th_{t+\tau}}{t} = \ep{\xi_t\xi_{t+\tau}}{t}$ for $\ep{h_th_{t+\tau}}{t}$, for any realization of $\xi_t = \bar{\xi}+\tld{\xi}_t$.
Averaging this equation over realizations of $\xi$ and using the DMFT equations, we get
\begin{equation}
  \label{eq:1-repl-ch}
  (1-\ddel_{\tau}^2)C^{\nam{h}}_{\tau} = I^2 + g^2(\bar{C}+\tld{C}_\tau) = I^2 + g^2 \ep{\phi\ev{h_t}\phi\ev{h_t}}{h_{t}\fl \oNor\ev{0,C^{\nam{h}}_{\tau}}},
\end{equation}
where $C^{\nam{h}}_{\tau}$ can be solved self-consistently for every choice of initial condition $C^{\nam{h}}_0, \lrpr{\ddel_\tau C^{\nam{h}}}_0$.

We empirically observe the order parameter is smooth at $\tau=0$, so we set $\lrpr{\ddel_\tau C^{\nam{h}}}_0 = 0$.
And for each choice of parameters $g$ and $I$, there is a unique initial position that is physical, i.e., satisfying the conditions $C^{\nam{h}}_\tau\leq C^{\nam{h}}_0$ (well defined covariance) and $C^{\nam{h}}_\infty$ exists (either chaos or time-independent)~\cite{crisanti2018Path}.
\eqref{eq:1-repl-ch} can be interpreted as a Newtonian system described by a potential defined up to $C^{\nam{h}}_0$, and the two conditions amount to requiring the potential at the boundary $C^{\nam{h}}_0$ to equal to its local maximum $C^{\nam{h}}_\infty$.
In general, $C^{\nam{h}}_{\infty}> I^2$, indicating induced quenched noise due to the coupling.
The potential can either be found directly using Price's theorem~\cite{clark2023Dimension}, or in the case of $\phi\ev{h} = \oerf\ev{\sqrt{\ppi}h/2}$ the force can be expressed as
\begin{equation}
  \label{eq:1-repl-ch-owen}
  \ddel_{\tau}^2C^{\nam{h}}_{\tau} =C^{\nam{h}}_{\tau} - I^2- g^2\Lrpr{1- \frac{4}{\ppi}\arctan\Ev{\sqrt{\frac{1+\frac{\ppi}{2}\lrpr{C^{\nam{h}}_{0}-C^{\nam{h}}_{\tau}}}{ 1+\frac{\ppi}{2}\lrpr{C^{\nam{h}}_{0}+C^{\nam{h}}_{\tau}}}}}}.
\end{equation}

Now, we numerically guess a value of $C^{\nam{h}}_0$, compute its corresponding potential over the domain (one might want to utilize Price's theorem), and compare the heights at the boundary and the local maximum.
If the local maximum is lower or does not exist, we increase the guess for $C^{\nam{h}}_0$, and if the boundary value is lower, we decrease the guess, until the two values are equal.
The trajectory of $C^{\nam{h}}_\tau$ can then be evolved numerically using any differential equation solver, under the Newtonian dynamics in \eqref{eq:1-repl-ch} with the chosen initial condition.
And the desired quantities $\bar{C}$ and $\tld{C}_\tau$ can be obtained simultaneously through the force according to \eqref{eq:1-repl-dmft-bar} and \eqref{eq:1-repl-dmft-tld}.
Details of the numeric procedure are in \methref{sec:ana-numerics}.

\subsection*{Fluctuation in autocovariance}\phantomsection\label{sec:dmft1-fluct}

In this section, we use the replica method to derive the fluctuation of $\tld{C}$ around its saddle-point value; a similar derivation can be found in \citep{clark2024Connectivity}.
Alternatively, this fluctuation can be obtained by following the cavity method in \citep{clark2023Dimension}.
Since we believe the network statistics are stationary, the variance of this fluctuation over realizations is the same as the variance $\ep{\tld{C}_{t,t+\infty}^2}{t}$ over time, used by \eqref{eq:pr-tld-result} and \eqref{eq:finite-pr-tld-result}.
Operationally, we assume that the autocovariance fluctuations inherit the time-translation invariance of the saddle-point, which is the approximation that allows the Fourier-space treatment below.
Like any saddle-point approximation, the covariance describing the fluctuation is the negative inverse of the Hessian of the action.
Since we separated the free energy into levels $F_{\bar{\nam{C}}}$, $F_{\tld{\nam{C}}}$, and $F_{\nam{h}}$ in \eqref{eq:1-repl-hierarchy} and we do not expect the dimensionality of temporal chaos to depend on the $\sim \pm 1/\sqrt{\nni}$ fluctuation of $\bar{C}$, we only need to calculate the Hessian of the action in $F_{\tld{\nam{C}}}$ in the middle of the hierarchy.
We will later show in subsubsection~\nameref{sec:full-hess} briefly how the results would be the same if we chose instead to not separate the levels and compute the full Hessian.

\subsubsection*{Hessian simplification and block-inversion}

Specifically, the Hessian has $2\times 2$ blocks corresponding to $\ddel_{\tld{C}_{t_1t_2}}$ and $\ddel_{\crc{\tld{C}}_{t_1t_2}}$, where the size of each block is $\frac{\nnt^2}{2}\times \frac{\nnt^2}{2}$:
\begin{equation}
  \label{eq:1-repl-hess}
  \begin{aligned}
    H_{t_1t_2,t'_1t'_2} &=\nni\bmx{ H\idc{^{\tld{\nam{C}}\tld{\nam{C}}}_{t_1t_2,t'_1t'_2}} &H\idc{^{\tld{\nam{C}}\crc{\tld{\nam{C}}}}_{t_1t_2,t'_1t'_2}} \\H\idc{^{\crc{\tld{\nam{C}}}\tld{\nam{C}}}_{t_1t_2,t'_1t'_2}} &H\idc{^{\crc{\tld{\nam{C}}}\crc{\tld{\nam{C}}}}_{t_1t_2,t'_1t'_2}}}
    \\ &= \nni\bmx{ \dd t^4 g^4 \ep{\crc{h}_{t_1}\crc{h}_{t_2},\crc{h}_{t'_1}\crc{h}_{t'_2}}{\xi} &-\ii\lrpr{\cchi_{t_1t'_1}\cchi_{t_2t'_2} +\dd t^2g^2 \ep{\crc{h}_{t_1}\crc{h}_{t_2},r_{t'_1}r_{t'_2}-\bar{r}^2}{\xi}} \\\cdots &-\ep{r_{t_1}r_{t_2}-\bar{r}^2, r_{t'_1}r_{t'_2}-\bar{r}^2}{\xi}}.
  \end{aligned}
\end{equation}
$\ep{f,g}{x}$ is a shorthand for the cumulant $\ep{fg}{x}-\ep{f}{x}\ep{g}{x}$, $\cchi_{t_2t'_2}$ here is the Kronecker delta, and the expression for $H^{\crc{\tld{\nam{C}}}\tld{\nam{C}}}$ is omitted since the Hessian is symmetric.
Using the trick $\ep{\crc{h}_{t}f}{\xi} = \ddel_{b_{t}}\ep{f}{\xi}/\lrpr{\ii\dd t}$, two of the blocks simplify to
\begin{equation}
  \label{eq:1-repl-hess-simp}
  \begin{aligned}
    &H\idc{^{\tld{\nam{C}}\tld{\nam{C}}}_{t_1t_2,t'_1t'_2}} = 0, \\&H\idc{^{\tld{\nam{C}}\crc{\tld{\nam{C}}}}_{t_1t_2,t'_1t'_2}} = -\ii\dd t^2 \lrpr{\ddelta_{t'_1-t_1}\ddelta_{t'_2-t_2} -g^2\ddel_{b_{t_1}}\ddel_{b_{t_2}}\ep{\tld{r}_{t'_1}\tld{r}_{t'_2}}{\xi}/\dd t^2},
  \end{aligned}
\end{equation}
where $\ddelta_{t'-t}$ here is the Dirac delta.
In this case, conveniently, we only need to invert the block $H\idc{^{\crc{\tld{\nam{C}}}\tld{\nam{C}}}^\inv}$ according to the Schur complement formulas
\begin{equation}
  \label{eq:1-repl-block-inv}
  H^\inv = \nni^{-1}\bmx{ -H\idc{^{\crc{\tld{\nam{C}}}\tld{\nam{C}}}^\inv}H\idc{^{\crc{\tld{\nam{C}}}\crc{\tld{\nam{C}}}}}H\idc{^{\crc{\tld{\nam{C}}}\tld{\nam{C}}}^\inv^\tsp} &H\idc{^{\crc{\tld{\nam{C}}}\tld{\nam{C}}}^\inv} \\\cdots & 0}.
\end{equation}

\subsubsection*{Multiplication and inversion in Fourier space in general}

The technical difficulty now is in performing the giant matrix multiplications and inversion, and the idea is to note that if two matrices are 2D-Toeplitz, then their matrix multiplication becomes scalar multiplication in Fourier space by the convolution theorem.
Specifically, if
\begin{equation}
  \label{eq:2d-toeplitz}
  A_{t_1t_2,t'_1t'_2} = \ofp_{\omega_1\omega_2}\ev{\crc{A}_{\omega_1\omega_2}}\lrpr{\ss{\ev{t'_1-t_1,t'_2-t_2}\\+\ev{t'_2-t_1,t'_1-t_2}}},
\end{equation}
meaning if $A$ is the sum of evaluating the Fourier transform of $\crc{A}$ at both $\lrpr{t'_1-t_1, t'_2-t_2}$ and $\lrpr{t'_2-t_1, t'_1-t_2}$, and similarly for $B$, then
\begin{equation}
  \label{eq:mat-ft-full}
  \sm{A_{t_1t_2,t'_1t'_2}B_{t'_1t'_2,t''_1t''_2}}{t'_1t'_2} =  2\cdot\Lrpr{\frac{\sqrt{2\ppi}}{\dd t}}^2\ofp_{\omega_1\omega_2}\ev{\crc{A}_{\omega_1\omega_2}\crc{B}_{\omega_1\omega_2}}\lrpr{\ss{\ev{t'_1-t_1,t'_2-t_2}\\+\ev{t'_2-t_1,t'_1-t_2}}}.
\end{equation}

The $\frac{\nnt^2}{2}\times \frac{\nnt^2}{2}$ Hessian blocks with $t_1\leq t_2$ are certainly not 2D-Toeplitz, but they would almost be 2D-Toeplitz if we symmetrize them by expanding them to $\nnt^2\times\nnt^2$ as $A_{t_1t_2,\cdots}=A_{t_2t_1,\cdots}$ if $t_1>t_2$.
In this case, the symmetrization of the product is $1/2$ the product of the symmetrizations, i.e.,
\begin{equation}
  \label{eq:mat-ft}
  \begin{aligned}
    \lrpr{AB}_{t_1t_2,t''_1t''_2} &=\sm{A_{t_1t_2,t'_1t'_2}B_{t'_1t'_2,t''_1t''_2}}{\ss{t'_1t'_2\\t'_1\leq t'_2}} =\frac{1}{2}\sm{A_{t_1t_2,t'_1t'_2}B_{t'_1t'_2,t''_1t''_2}}{t'_1t'_2}
    \\&= \frac{2\ppi}{\dd t^2}\ofp_{\omega_1\omega_2}\ev{\crc{A}_{\omega_1\omega_2}\crc{B}_{\omega_1\omega_2}}\lrpr{\ss{\ev{t'_1-t_1,t'_2-t_2}\\+\ev{t'_2-t_1,t'_1-t_2}}},
  \end{aligned}
\end{equation}
where $\crc{A}$ denotes the 2D Fourier transform of $A$ and likewise for $\crc{B}$, and the Fourier transforms of the symmetrizations are usually easy to obtain.

Finally, we comment on how to deal with the fact that the Hessian blocks are not exactly 2D-Toeplitz but a 2D-Toeplitz part $A$ plus a deviation $\XX$ that is $\sim 1$ at certain entries.
Since we are only interested in the particular variance $\ep{\tld{C}_{t,t+\infty}^2}{t}$ of the fluctuations, as long as $\XX$ does not contribute in the limit $t_2-t_1,t'_2-t'_1\to\infty$ of interest, we can safely ignore them.
Conveniently, we will see that for all blocks, $\XX$ itself is exponentially suppressed by either $t_2-t_1$ or $t'_2-t'_1$.
Then, their contributions through multiplication and inversion given by
\begin{equation}
  \label{eq:no-contribution}
  \begin{gathered}
    \lrpr{A + \XX}\lrpr{B + \XX} = AB + A\XX+\XX B+\XX\XX
    \\\lrpr{A + \XX}^\inv = A^\inv - A^\inv \XX\lrpr{A + \XX}^\inv.
  \end{gathered}
\end{equation}
would also vanish in the limit of interest, as long as $A$, $B$, and $A^\inv$ are 2D-Toeplitz (or $\XX$-like) and do not grow exponentially with any time difference (and since inverses are unique).
We will show that this is also true, since their Fourier transforms exist.

\subsubsection*{Explicit inversion of the Hessian}

We now use the general method to find the Hessian's inverse in \eqref{eq:1-repl-block-inv}.
Here and below, $r'_t\equiv\phi'\ev{h_t}$ and $r''_t\equiv\phi''\ev{h_t}$.
Using the trick $\ep{\crc{h}_{t}f}{\xi} =\ddel_{b_{t}}\ep{f}{\xi}/\lrpr{\ii\dd t}$, the expectation in $H\idc{^{\crc{\tld{\nam{C}}}\tld{\nam{C}}}}$ becomes
\begin{equation}
  \label{eq:1-repl-crc-non}
  \begin{aligned}
    \ddel_{b_{t_1}}\ddel_{b_{t_2}}\ep{\tld{r}_{t'_1}\tld{r}_{t'_2}}{\xi}/\dd t^2 =& \green_{t'_1-t_1}\green_{t'_2-t_2}\ep{r'_{t'_1}r'_{t'_2}}{\xi}+\green_{t'_1-t_2}\green_{t'_2-t_1}\ep{r'_{t'_1}r'_{t'_2}}{\xi} \\&+\green_{t'_1-t_1}\green_{t'_1-t_2}\ep{r''_{t'_1}r_{t'_2}}{\xi} +\green_{t'_2-t_1}\green_{t'_2-t_2}\ep{r_{t'_1}r''_{t'_2}}{\xi}
    \\= & \green_{t'_1-t_1}\green_{t'_2-t_2}\ep{\ep{r'_t}{\tld{\xi}_t}^2}{\bar{\xi}} +\green_{t'_1-t_2}\green_{t'_2-t_1}\ep{\ep{r'_t}{\tld{\xi}_t}^2}{\bar{\xi}} +\XX_{t_1t_2|| t'_1t'_2},
  \end{aligned}
\end{equation}
where $\green_{\tau}=\TTheta\ev{\tau}\ee^{-\tau}$ is the Green's function of the operator $\lrpr{1+\ddel_\tau}$ in the 1D Langevin dynamics.
One can check that indeed the two terms outside of $\XX$ are 2D-Toeplitz, and the remaining $\XX$ vanishes with $t_2-t_1$ or $t'_2-t'_1$.
To track which pairings cause $\XX$ to vanish, we put them in its subscript.
In addition to the expectation, $H\idc{^{\crc{\tld{\nam{C}}}\tld{\nam{C}}}}$ has a Dirac delta term which is also 2D-Toeplitz after symmetrization, and we collect all 2D-Toeplitz terms in $A$ for brevity and write $H\idc{^{\crc{\tld{\nam{C}}}\tld{\nam{C}}}} = -\ii\dd t^2\lrpr{A + \XX}$.

Using \eqref{eq:mat-ft}, the equation $\sm{A_{t_1t_2,t'_1t'_2}A\idc{^{\inv}_{t'_1t'_2,t''_1t''_2}}}{ \ss{t'_1t'_2\\t'_1\leq t'_2}} = \cchi_{t_1t''_1}\cchi_{t_2t''_2}$ for inversion becomes
\begin{equation}
  \label{eq:1-repl-mat-ft-inv}
  \begin{gathered}
    \frac{1}{2}\sm{A_{t_1t_2,t'_1t'_2}A\idc{^{\inv}_{t'_1t'_2,t''_1t''_2}}}{t'_1t'_2} = \lrpr{\cchi_{t_1t''_1}\cchi_{t_2t''_2} + \cchi_{t_1t''_2}\cchi_{t_2t''_1}}\\
    \frac{2\ppi}{\dd t^2}\crc{A}_{\omega_1\omega_2}\crc{A}^{\nam{\inv}}_{\omega_1\omega_2} = \frac{\dd t^2}{2\ppi}
  \end{gathered}
\end{equation}
after symmetrization, and the Fourier space version in the second line assumes $A^\inv$ is also 2D-Toeplitz.
So by finding $\crc{A}$ for $A$ according to \eqref{eq:2d-toeplitz},
\begin{equation}
  \label{eq:1-repl-crc-non-inv}
  H\idc{^{\crc{\tld{\nam{C}}}\tld{\nam{C}}}^\inv_{t_1t_2,t'_1t'_2}} = \ofp_{\omega_1\omega_2}\Ev{-\frac{\dd t^2}{2\ppi\ii\lrpr{1 -2\ppi g^2 \ep{\ep{r'_t}{\tld{\xi}_t}^2}{\bar{\xi}}\crc{\green}_{\omega_1}\crc{\green}_{\omega_2}}}}\lrpr{\ss{\ev{t'_1-t_1,t'_2-t_2}\\+\ev{t'_2-t_1,t'_1-t_2}}} + \XX_{t_1t_2|| t'_1t'_2},
\end{equation}
where $\XX$ here is different from before but still vanishes with the time difference as paired.

$H\idc{^{\crc{\tld{\nam{C}}}\tld{\nam{C}}}^\inv^\tsp}$ is the transpose of $H\idc{^{\crc{\tld{\nam{C}}}\tld{\nam{C}}}^\inv}$, so they are the same up to complex conjugating $\crc{\green}_\omega$-s, and the only other block that needs calculation according to \eqref{eq:1-repl-block-inv} is $H\idc{^{\crc{\tld{\nam{C}}}\crc{\tld{\nam{C}}}}}$.
Since $\tld{r}_{t}$ is a sum over $\nni\gg 1$ weakly correlated contributions for a given quenched noise $\bar{\xi}$, we approximate it as Gaussian by a central-limit argument (its variance varies over $\bar{\xi}$, so it is very non-Gaussian overall).
Wick's theorem then gives
\begin{equation}
  \label{eq:1-repl-crc-crc}
  \begin{aligned}
    -H\idc{^{\crc{\tld{\nam{C}}}\crc{\tld{\nam{C}}}}} =&\ep{\ep{r_{t_1}r_{t'_1}}{\tld{\xi}}\ep{r_{t_2}r_{t'_2}}{\tld{\xi}} -\bar{r}^4}{\bar{\xi}} +\ep{\ep{r_{t_1}r_{t'_2}}{\tld{\xi}}\ep{r_{t_2}r_{t'_1}}{\tld{\xi}} -\bar{r}^4}{\bar{\xi}} +\XX_{t_1t_2|| t'_1t'_2}
    \\=&\ofp_{\omega_1\omega_2}\ev{\crc{\tld{K}}_{\omega_1\omega_2}}\lrpr{ \ss{\ev{t'_1-t_1,t'_2-t_2}\\+\ev{t'_2-t_1,t'_1-t_2}}} +\XX_{t_1t_2|| t'_1t'_2},
  \end{aligned}
\end{equation}
where $\crc{\tld{K}}$ is the Fourier transform of $\tld{K}_{\tau_1\tau_2} =\ep{\ep{r_{t}r_{t+\tau_1}}{\tld{\xi}}\ep{r_{t}r_{t+\tau_2}}{\tld{\xi}} -\bar{r}^4}{\bar{\xi}}$.
So again using \eqref{eq:mat-ft}, the covariance of $\tld{C}$ over realizations is the $\tld{\nam{C}}\tld{\nam{C}}$ (upper left) block of $-H^\inv$
\begin{equation}
  \label{eq:1-repl-inv-non-non}
   \ep{\tld{C}_{t_1t_2}\tld{C}_{t'_1t'_2}}{\dbl{\tld{C}}} =\frac{1}{\nni} \ofp_{\omega_1\omega_2}\Ev{\frac{\crc{\tld{K}}_{\omega_1\omega_2}}{\lrvt{1 -2\ppi g^2 \ep{\ep{r'_t}{\tld{\xi}_t}^2}{\bar{\xi}}\crc{\green}_{\omega_1}\crc{\green}_{\omega_2}}^2}}\lrpr{\ss{\ev{t'_1-t_1,t'_2-t_2}\\+\ev{t'_2-t_1,t'_1-t_2}}} +\XX_{t_1t_2|| t'_1t'_2}.
\end{equation}

We now evaluate the expectations explicitly so that we can obtain a numerical value for $\ep{\tld{C}_{t,t+\infty}^2}{\dbl{\tld{C}}}$ to be used in \eqref{eq:pr-tld-result} and \eqref{eq:finite-pr-tld-result}.
For $\phi\ev{h} = \oerf\ev{\sqrt{\ppi}h/2}$, in terms of $C^{\nam{h}}$ and $\tld{C}$ in \eqref{eq:1-repl-ch-owen} solved at the saddle-point, the expectations are given by
\begin{equation}
  \label{eq:1-repl-fluct-owen}
  \begin{gathered}
    \ep{\ep{r'_t}{\tld{\xi}_t}^2}{\bar{\xi}} = \frac{1}{ \sqrt{\lrpr{1+\frac{\ppi}{2}\lrpr{C^{\nam{h}}_{0}-C^{\nam{h}}_{\infty}}} \lrpr{1+\frac{\ppi}{2}\lrpr{C^{\nam{h}}_{0}+C^{\nam{h}}_{\infty}}}}},\\
    \begin{aligned}
      K_{\tau_1\tau_2} =&\ep{\ep{r_{t}r_{t+\tau_1}}{\tld{\xi}}\ep{r_{t}r_{t+\tau_2}}{\tld{\xi}}}{\bar{\xi}}
      \\=& 1 - \frac{4}{\ppi}\Sm{\arctan\Ev{ \sqrt{\frac{1+\frac{\ppi}{2}\lrpr{C^{\nam{h}}_{0}-C^{\nam{h}}_{\tau}}}{ 1+\frac{\ppi}{2}\lrpr{C^{\nam{h}}_{0}+C^{\nam{h}}_{\tau}}}}}}{\tau=\tau_1,\tau_2}
      \\ &+64\int_0^{c_1}\int_0^{c_2}\dd y\dd z \frac{1}{\lrpr{2\ppi}^2}\frac{1}{\lrpr{1+y^2}\lrpr{1+z^2}}\frac{1}{\sqrt{\alpha b+1}},
    \end{aligned}
  \end{gathered}
\end{equation}
where $\alpha=2+y^2+z^2$, $b = \frac{\frac{\ppi}{2}C^{\nam{h}}_{\infty}}{ 1 +\frac{\ppi}{2}\lrpr{C^{\nam{h}}_{0} -C^{\nam{h}}_\infty}}$, and
$c_{i=1,2} = \sqrt{\frac{1 +\frac{\ppi}{2} \lrpr{C^{\nam{h}}_{0}-C^{\nam{h}}_{\tau_i}}}{ 1 +\frac{\ppi}{2} \lrpr{C^{\nam{h}}_{0}+C^{\nam{h}}_{\tau_i}-2C^{\nam{h}}_\infty}}}$.
$\tld{K}$ is then $\tld{K}_{\tau_1\tau_2} = K_{\tau_1\tau_2} - K_{\infty\infty}$.
We note that in this problem $\tld{K}_{\tau_1\tau_2} \neq\ep{\ep{\tld{r}_{t}\tld{r}_{t+\tau_1}}{\tld{\xi}} \ep{\tld{r}_{t}\tld{r}_{t+\tau_2}}{\tld{\xi}}}{\bar{\xi}}$.
Alternatively, $K$ can be directly computed numerically using its definition, but that is less efficient, since it would involve an integral over infinite support, while the expression in \eqref{eq:1-repl-fluct-owen} does not.
We compute the Fourier transform $\crc{K}$ and the inverse Fourier transform in \eqref{eq:1-repl-inv-non-non} numerically, detailed in \methref{sec:ana-numerics}.

\subsubsection*{Equivalence to inverting full Hessian}\phantomsection\label{sec:full-hess}

We now show the equivalence between inverting the Hessian for $F_{\bar{\nam{C}}}$ and $F_{\tld{\nam{C}}}$ level by level and inverting the Hessian for the full $F$.
In the second case, the full Hessian
\begin{equation}
  \label{eq:1-repl-full-hess}
  H^{\nam{f}} =\nni\sbmx{ 0&H^{\bar{\nam{C}}\crc{\bar{\nam{C}}}}&0&H^{\bar{\nam{C}}\crc{\tld{\nam{C}}}} \\\cdots&H^{\crc{\bar{\nam{C}}}\crc{\bar{\nam{C}}}}&0&H^{\crc{\bar{\nam{C}}}\crc{\tld{\nam{C}}}} \\\cdots&\cdots&0&H^{\tld{\nam{C}}\crc{\tld{\nam{C}}}} \\\cdots&\cdots&\cdots&H^{\crc{\tld{\nam{C}}}\crc{\tld{\nam{C}}}}}
\end{equation}
would instead have $4\times 4$ blocks, and the $2\times 2$ sub-block for $\dbl{\tld{C}}$ in the lower right would be the same as in \eqref{eq:1-repl-hess}.
The zeros mostly come from expectations of $\crc{h}$ only, and one of them is the susceptibility of $\bar{r}$ to a local Dirac delta perturbation.
The fluctuations in $\dbl{\bar{C}}$ and $\dbl{\tld{C}}$ are coupled through the two off-diagonal $2\times 2$ blocks, where their sub-blocks are $1\times\nnt^2/2$ coupling $t_1$ to $t_2$.
Both sub-blocks are exponentially suppressed by $t_2-t_1$:
\begin{equation}
  \label{eq:1-repl-full-xx}
  \begin{aligned}
    &H^{\bar{\nam{C}}\crc{\tld{\nam{C}}}}_{t_2-t_1} \propto\ep{\ep{r'_{t_1}r'_{t_2}}{\tld{\xi}} +\ep{r''_{t_1}r_{t_2}}{\tld{\xi}}}{\bar{\xi}} -\ep{\ep{r'_t}{\tld{\xi}_t}^2 +\ep{r''_t}{\tld{\xi}_t}\ep{r_t}{\tld{\xi}_t}}{\bar{\xi}},
    \\ &H^{\crc{\bar{\nam{C}}}\crc{\tld{\nam{C}}}}_{t_2-t_1} \propto\ep{\bar{r}^2\ep{\tld{r}_{t_1}\tld{r}_{t_2}}{\tld{\xi}}}{\bar{\xi}} -\ep{\bar{r}^2}{\bar{\xi}}\ep{\tld{r}_{t_1}\tld{r}_{t_2}}{\xi},
  \end{aligned}
\end{equation}
$H^{\bar{\nam{C}}\crc{\tld{\nam{C}}}}$ vanishes by cancellation, and terms in $H^{\crc{\bar{\nam{C}}}\crc{\tld{\nam{C}}}}$ vanish independently.

By the Schur complement formulas, the covariance for $\dbl{\tld{C}}$ is this time the negative inverse of the effective Hessian
\begin{equation}
  \label{eq:1-repl-full-schur}
  H^{\nam{e}} = H + \sbmx{0&0\\0&\frac{H^{\crc{\bar{\nam{C}}}\crc{\bar{\nam{C}}}}}{\lrpr{H^{\bar{\nam{C}}\crc{\bar{\nam{C}}}}}^2}H^{\bar{\nam{C}}\crc{\tld{\nam{C}}}\tsp}H^{\bar{\nam{C}}\crc{\tld{\nam{C}}}} -\frac{1}{H^{\bar{\nam{C}}\crc{\bar{\nam{C}}}}}\lrpr{H^{\crc{\bar{\nam{C}}}\crc{\tld{\nam{C}}}\tsp}H^{\bar{\nam{C}}\crc{\tld{\nam{C}}}}+H^{\bar{\nam{C}}\crc{\tld{\nam{C}}}\tsp}H^{\crc{\bar{\nam{C}}}\crc{\tld{\nam{C}}}}}},
\end{equation}
where every correction term to the old Hessian in \eqref{eq:1-repl-hess} is a product of $H^{\bar{\nam{C}}\crc{\tld{\nam{C}}}}$ and $H^{\crc{\bar{\nam{C}}}\crc{\tld{\nam{C}}}}$ and therefore ends up in $\XX$.
So our intuitive separation between $F_{\bar{\nam{C}}}$ and $F_{\tld{\nam{C}}}$ is valid.

\subsection*{Two replicas for two external inputs}\phantomsection\label{sec:dmft2}

We now consider the same procedure used so far but for two replicas of the same network, sharing the same coupling $J$ but driven by distinct (and correlated) external inputs $f_1$ and $f_2$.
We index the two replicas by $a = 1, 2$, and recall that the correlation between the two external inputs is described by $\spmx{f_{1,i}\\f_{2,i}}\fl\oNor\ev{0,I^2\rho_{a_1a_2}}$ for every neuron $i$, and $\rho=\spmx{1&\rhoin\\ \rhoin&1}$ is the correlation matrix.
Our goal is to find the value and fluctuation of the new statistics $\bar{C}_{12}$ and $\tld{C}_{12\tau}$, used in \eqref{eq:cs},  \eqref{eq:os-tld-result}, and \eqref{eq:pr-bar-result}.

\subsubsection*{New statistics and their saddle-point values}

We follow the same procedure detailed in the previous two subsections, so we only describe what appears different.
Starting from the quenched-disorder-dependent partition function for both replicas, the free energy assuming replica symmetry (symmetric across replica pairs) is given by
\begin{equation}
  \label{eq:2-repl-zq-f}
  \begin{aligned}
    Z_q &=\int \pd{\dd h_{a,i,t}}{ait}\Pd{\ddelta\Ev{\dd t \Lrpr{\lrpr{h_{a,i,t}+\ddel_th_{a,i,t}} -\sm{J_{ij}\phi\ev{h_{a,j,t-\dd t}}}{j} -f_{a,i} -b_{a,i,t}}}}{ait} \ee^{\ii\sm{\nt{\crc{b}_{a,i,t}h_{a,i,t}}{t}}{ai}}
    \\ F&= \log \int \Pd{\frac{\dd h_{a,i,t}\dd \crc{h}_{a,i,t}}{2\pi}}{ait}\prod_{i}\Lbk\exp\Ev{\ii\Sm{\Nt{\mtx{-\crc{h}_{a,i,t}\lrpr{h_{a,i,t}+\ddel_th_{a,i,t}}\\+\crc{h}_{a,i,t}b_{a,i,t}+\crc{b}_{a,i,t}h_{a,i,t}}}{t}}{a}}\\&\qquad\times\exp\Ev{-\frac{1}{2}\Sm{\Nt{\crc{h}_{a_1,i,t_1}\crc{h}_{a_2,i,t_2}\Lrpr{g^2\Sm{\frac{1}{\nni}\phi\ev{h_{a_1,j,t_1}}\phi\ev{h_{a_2,j,t_2}}}{j}+I^2\rho_{a_1a_2}}}{t_1t_2}}{a_1a_2}}\Rbk,
  \end{aligned}
\end{equation}
and the new replica index $a$ behaves similarly to the time index $t$ (the neuron index $i$ is special since it is the index over which the quenched disorder is independent).
Like in \eqref{eq:1-repl-f}, the factor $\sm{\phi\ev{h_{a_1,j,t_1}}\phi\ev{h_{a_2,j,t_2}}}{j}/\nni$ that couples different neurons is exactly the statistics of interest, so we introduce $\bar{C}_{a_1a_2} = \sm{\ep{\phi\ev{h_{a_1,i,t}}}{t}\ep{\phi\ev{h_{a_2,i,t}}}{t}}{i}/\nni$ and $\tld{C}_{a_1a_2t_1t_2} = \sm{\phi\ev{h_{a_1,i,t_1}}\phi\ev{h_{a_2,i,t_2}}-\ep{\phi\ev{h_{a_1,i,t}}}{t}\ep{\phi\ev{h_{a_2,i,t}}}{t}}{i}/\nni$ as variables using $\ddelta$-s.
Again due to symmetry, $\bar{C}$ has $3$ degrees of freedom, where the two same-replica values $\bar{C}_{11}$ and $\bar{C}_{22}$ should be statistically equivalent, and $\bar{C}_{12}$ is the only independent cross-replica degree.
Similarly, $\tld{C}$ has $(2\nnt)^2/2$ degrees of freedom, including two equivalent collections of same-replica values, where each collection contains $\nnt^2/2$ degrees, and one collection of cross-replica values with $\nnt^2$ degrees ($C_{12,t_1t_2}\neq C_{12,t_2t_1}$).

Again, the dynamics can be fully decoupled over neurons $i$ by assuming the perturbations $b$ and currents $\crc{b}$ are spatially uniform, and the hierarchy of free energies is
almost exactly the same as that in \eqref{eq:1-repl-hierarchy}, except for the additional sums over $a$:
\begin{equation}
  \label{eq:2-repl-hierarchy}
  \begin{aligned}
    F_{\bar{\nam{C}}} =& \log \int\Pd{\frac{\dd\bar{C}_{a_1a_2}\dd\crc{\bar{C}}_{a_1a_2}}{2\pi/\nni}}{ \ss{a_1a_2\\a_1\leq a_2}} \exp\ev{-\ii \nni\sm{\crc{\bar{C}}_{a_1a_2}\bar{C}_{a_1a_2}}{ \ss{a_1a_2\\a_1\leq a_2}} +F_{\tld{\nam{C}}}},
    \\F_{\tld{\nam{C}}} =&\log\int\Pd{\frac{\dd \tld{C}_{a_1a_2t_1t_2} \dd\crc{\tld{C}}_{a_1a_2t_1t_2}}{2\pi/\nni}}{ \ss{a_1a_2t_1t_2\\ a_1,t_1\leq a_2,t_2}}\exp\ev{-\ii \nni\sm{\crc{\tld{C}}_{a_1a_2t_1t_2}\tld{C}_{a_1a_2t_1t_2}}{ \ss{a_1a_2t_1t_2\\ a_1,t_1\leq a_2,t_2}}+\nni F_{\nam{h}}},
    \\F_{\nam{h}} =& \log\int \Pd{\frac{\dd h_{at}\dd \crc{h}_{at}}{2\pi}}{at} \exp\Ev{\ii\sm{\crc{\tld{C}}_{a_1a_2t_1t_2}\lrpr{r_{a_1t_1}r_{a_2t_2} - \bar{r}_{a_1}\bar{r}_{a_2}}}{ \ss{a_1a_2t_1t_2\\ a_1,t_1\leq a_2,t_2}}+\ii\sm{\crc{\bar{C}}_{a_1a_2}\bar{r}_{a_1}\bar{r}_{a_2}}{\ss{a_1a_2\\a_1\leq a_2}}} \\&\qquad\qquad\times\exp\Ev{\ii\Sm{\Nt{\mtx{-\crc{h}_{at}\lrpr{h_{at}+\ddel_th_{at}}\\+\crc{h}_{at}b_{at}+\crc{b}_{at}h_{at}}}{t}}{a}} \\&\qquad\qquad\times\exp\Ev{-\frac{1}{2}\Sm{\Nt{\crc{h}_{a_1t_1}\crc{h}_{a_2t_2}\Lrpr{g^2\lrpr{\bar{C}_{a_1a_2}+\tld{C}_{a_1a_2t_1t_2}}+I^2\rho_{a_1a_2}}}{t_1t_2}}{a_1a_2}}.
  \end{aligned}
\end{equation}
We can again perform saddle-point approximations to the first two levels, and in addition to the old DMFT equations in \eqref{eq:1-repl-dmft-bar} and \eqref{eq:1-repl-dmft-tld} for each of the two replicas, we also get
\begin{equation}
  \label{eq:2-repl-dmft}
  \begin{aligned}
    &\ii\crc{\bar{C}}_{12}= -g^2 \nt{\ep{\ep{\crc{h}_{1t_1}\crc{h}_{2t_2}}{\dbl{h}}}{\dbl{\tld{C}}}}{t_1t_2},
    \qquad&& \ii\crc{\tld{C}}_{12t_1t_2} =-\dd t^2 g^2\ep{\crc{h}_{1t_1}\crc{h}_{2t_2}}{\dbl{h}},
    \\ &\bar{C}_{12} =\ep{\ep{\bar{r}_{1}\bar{r}_{2}}{\dbl{h}}}{\dbl{\tld{C}}},
    \qquad&&\tld{C}_{12t_1t_2} = \ep{r_{1t_1}r_{2t_2}-\bar{r}_1\bar{r}_2}{\dbl{h}}.
  \end{aligned}
\end{equation}
To evaluate the expectations over $\dbl{h}$, we again rewrite the last level $F_{\nam{h}}$ as the free energy of an equivalent Langevin dynamics, this time in 2D for the two replicas:
\begin{equation}
  \label{eq:2-repl-equiv-dyn}
  \begin{aligned}
    F_{\nam{h}} =&\log\int \Pd{\frac{\dd h_{at}\dd \crc{h}_{at}}{2\pi}}{at} \exp\Ev{\ii\sm{\crc{\tld{C}}_{a_1a_2t_1t_2}\lrpr{r_{a_1t_1}r_{a_2t_2} - \bar{r}_{a_1}\bar{r}_{a_2}}}{ \ss{a_1a_2t_1t_2\\ a_1,t_1\leq a_2,t_2}}+\ii\sm{\crc{\bar{C}}_{a_1a_2}\bar{r}_{a_1}\bar{r}_{a_2}}{\ss{a_1a_2\\a_1\leq a_2}}}
    \\ &\qquad\times\Ep{\prod_a \Ep{\exp\Ev{\ii\Nt{\mtx{-\crc{h}_{at}\lrpr{h_{at}+\ddel_th_{at}}\\ +\crc{h}_{at}b_{at} +\crc{b}_{at}h_{at}}}{t} +\ii\Nt{\crc{h}_{at}\lrpr{\bar{\xi}_a +\tld{\xi}_{at}}}{t}}}{\tld{\xi}_a}}{\bar{\xi}}.
  \end{aligned}
\end{equation}
The two Langevin dynamics are only coupled by having correlated noise, but otherwise they evolve independently over time.
The quenched noises are certainly correlated across replicas as $\spmx{\bar{\xi}_1\\ \bar{\xi}_2}\fl \oNor\ev{0,g^2\bar{C}_{a_1a_2}+I^2\rho_{a_1a_2}}$.
Correlated cross-replica thermal noise would correspond to a different self-consistent solution, and when the two replicas share the same initial condition with $\rhoin=1$, that solution is related to the system's maximum Lyapunov exponent~\cite{crisanti2018Path}.
For the present observables, we use the saddle point where $\tld{C}_{12t_1t_2}=0$, so $\tld{\xi}_1$ and $\tld{\xi}_2$ are independent by construction in \eqref{eq:2-repl-equiv-dyn}.
Using \eqref{eq:2-repl-dmft}, we can see that $\crc{\bar{C}}=\crc{\tld{C}}=0$ and $\tld{C}_{12}=0$ are self-consistent.

Since each Langevin dynamics evolves independently, the saddle-point values of the same-replica quantities are equal to their single-replica versions calculated in \eqref{eq:1-repl-ch-owen}, i.e., $\bar{C}_{aa}=\bar{C}$, $\tld{C}_{aa\tau}=\tld{C}_\tau$, and $C^{\nam{h}}_{aa\tau} = C^{\nam{h}}_{\tau}$, and it only remains to solve for $\bar{C}_{12}$.
For $\phi\ev{h} = \oerf\ev{\sqrt{\ppi}h/2}$, \eqref{eq:2-repl-dmft} gives the algebraic equation
\begin{equation}
  \label{eq:2-repl-ch-owen}
  \bar{C}_{12} =\ep{\ep{\phi\ev{h_{1t}}}{\tld{\xi}}\ep{\phi\ev{h_{2t}}}{\tld{\xi}}}{\bar{\xi}} =\Lrpr{1-\frac{4}{\ppi}\arctan\Ev{\sqrt{\frac{ 1+\frac{\ppi}{2}\lrpr{C^{\nam{h}}_{0}-\lrpr{g^2\bar{C}_{12}+I^2\rhoin}}}{ 1+\frac{\ppi}{2}\lrpr{C^{\nam{h}}_{0}+\lrpr{g^2\bar{C}_{12}+I^2\rhoin}}}}}}
\end{equation}
in terms of the single-replica $C^{\nam{h}}_{0}$, using Eqs.~(20.010.8) and (10.010.8) of \citep{owen1980Table}.
Note that when $\rhoin<0$, the argument of $\arctan$ could be $>1$, so $\bar{C}_{12}<0$ as expected.
By solving \eqref{eq:2-repl-ch-owen} numerically, we can find a numerical value for $\oCS$ in \eqref{eq:cs}.

\subsubsection*{Fluctuations in cross-replica order parameters}

We now move to the fluctuations in $\dbl{\tld{C}}_{12}$ and $\dbl{\bar{C}}_{12}$.
Since the argument in \nameref{sec:full-hess} still holds, we can find the two fluctuations independently.

On the level of $\dbl{\tld{C}}$, before writing down the Hessian, we first consider which entries in it would vanish.
Since the two Langevin dynamics evolve independently and the two thermal noises are assumed to be independent, $H^{\tld{\nam{C}}_{11}\tld{\nam{C}}_{22}} =H^{\crc{\tld{\nam{C}}}_{11}\tld{\nam{C}}_{22}} =H^{\tld{\nam{C}}_{11}\crc{\tld{\nam{C}}}_{22}} =0$ and $H^{\crc{\tld{\nam{C}}}_{11}\crc{\tld{\nam{C}}}_{22}}=\XX$.
Similarly, we can find that $H^{\dbl{\tld{\nam{C}}}_{11}\tld{\nam{C}}_{12}} =H^{\dbl{\tld{\nam{C}}}_{22}\tld{\nam{C}}_{12}}=0$ and $H^{\dbl{\tld{\nam{C}}}_{11}\crc{\tld{\nam{C}}}_{12}}, H^{\dbl{\tld{\nam{C}}}_{22}\crc{\tld{\nam{C}}}_{12}}=\XX$.
This means if we order the blocks as $\ddel_{\tld{C}_{11}}, \ddel_{\crc{\tld{C}}_{11}}, \ddel_{\tld{C}_{22}}, \ddel_{\crc{\tld{C}}_{22}}, \ddel_{\tld{C}_{12}}, \ddel_{\crc{\tld{C}}_{12}}$,
\begin{equation}
  \label{eq:2-repl-tld-hess}
  H = \sbmx{H^{\dbl{\tld{\nam{C}}}_{11}\dbl{\tld{\nam{C}}}_{11}}&\smtx{0&0\\0&\XX}&\smtx{0&\XX\\0&\XX}
    \\\cdots&H^{\dbl{\tld{\nam{C}}}_{22}\dbl{\tld{\nam{C}}}_{22}}&\smtx{0&\XX\\0&\XX}
    \\\cdots&\cdots&H^{\dbl{\tld{\nam{C}}}_{12}\dbl{\tld{\nam{C}}}_{12}}},
\end{equation}
and the three blocks $H^{\dbl{\tld{\nam{C}}}_{11}\dbl{\tld{\nam{C}}}_{11}}$, $H^{\dbl{\tld{\nam{C}}}_{22}\dbl{\tld{\nam{C}}}_{22}}$, and $H^{\dbl{\tld{\nam{C}}}_{12}\dbl{\tld{\nam{C}}}_{12}}$ can be inverted independently up to corrections from $\XX$.

For fluctuations in $\dbl{\tld{C}}_{12}$, we only need to invert the cross-replica Hessian
\begin{equation}
  \label{eq:2-repl-tld-hess-12}
  \begin{aligned}
    H^{\dbl{\tld{\nam{C}}}_{12}\dbl{\tld{\nam{C}}}_{12}}_{t_1t_2,t'_1t'_2} &= \nni\bmx{0 &-\ii\lrpr{\cchi_{t_1t'_1}\cchi_{t_2t'_2} +\dd t^2g^2 \ep{\crc{h}_{1t_1}\crc{h}_{2t_2},r_{1t'_1}r_{2t'_2}-\bar{r}_1\bar{r}_2}{\xi}} \\\cdots &-\ep{r_{1t_1}r_{2t_2}-\bar{r}_1\bar{r}_2, r_{1t'_1}r_{2t'_2}-\bar{r}_1\bar{r}_2}{\xi}},
  \end{aligned}
\end{equation}
which has the same block structure as the same-replica Hessian in \eqref{eq:1-repl-hess-simp}, so \eqref{eq:1-repl-block-inv} still applies.
But this cross-replica Hessian is much simpler to invert, since the blocks here are exactly 2D-Toeplitz depending only on the pairing $t'_1-t_1$ and $t'_2-t_2$, and are already $\nnt^2\times\nnt^2$.
So \eqref{eq:mat-ft-full} is unnecessary, and the convolution theorem can be applied directly, yielding
\begin{equation}
  \label{eq:2-repl-tld-inv-non-non}
  \begin{aligned}
    \ep{\tld{C}_{12t_1t_2}\tld{C}_{12t'_1t'_2}}{\dbl{\tld{C}}} =&\frac{1}{\nni} \ofp_{\omega_1\omega_2}\Ev{\frac{\crc{\tld{K}}_{12\omega_1\omega_2}}{\lrvt{1 -2\ppi g^2 \ep{\ep{r'_{1t}}{\tld{\xi}_1}\ep{r'_{2t}}{\tld{\xi}_2}}{\bar{\xi}}\crc{\green}_{\omega_1}\crc{\green}_{\omega_2}}^2}}\ev{t'_1-t_1,t'_2-t_2}
    \\ &+\XX_{t_1t_2|| t'_1t'_2}.
  \end{aligned}
\end{equation}

When $\phi\ev{h} = \oerf\ev{\sqrt{\ppi}h/2}$, in terms of the single-replica $C^{\nam{h}}$ and $\tld{C}$ in \eqref{eq:1-repl-ch-owen} and the cross-replica $\bar{C}_{12}$ in \eqref{eq:2-repl-ch-owen}, the numerical values of the expectations are
\begin{equation}
  \label{eq:2-repl-tld-fluct-owen}
  \begin{gathered}
    \ep{\ep{r'_{1t}}{\tld{\xi}_1}\ep{r'_{2t}}{\tld{\xi}_2}}{\bar{\xi}} =\frac{1}{\sqrt{\lrbk{1+\frac{\ppi}{2}\lrpr{C^{\nam{h}}_{0} -\lrvt{g^2\bar{C}_{12}+I^2\rhoin}}} \lrbk{1+\frac{\ppi}{2}\lrpr{C^{\nam{h}}_{0} +\lrvt{g^2\bar{C}_{12}+I^2\rhoin}}}}},\\
    \begin{aligned}
      K_{12\tau_1\tau_2} =&\ep{\ep{r_{1t}r_{1,t+\tau_1}}{\tld{\xi}_1}\ep{r_{2,t}r_{2,t+\tau_2}}{\tld{\xi}_2}}{\bar{\xi}}
      \\=& 1 - \frac{4}{\ppi}\Sm{\arctan\Ev{ \sqrt{\frac{1+\frac{\ppi}{2} \lrpr{C^{\nam{h}}_{0} -\lrvt{g^2\bar{C}_{12}+I^2\rhoin}}}{ 1+\frac{\ppi}{2} \lrpr{C^{\nam{h}}_{0} +\lrvt{g^2\bar{C}_{12}+I^2\rhoin}}}}}}{\tau=\tau_1,\tau_2}
      \\ &+64\int_0^{c_1}\int_0^{c_2}\dd y\dd z \frac{1}{\lrpr{2\ppi}^2}\frac{1}{\lrpr{1+y^2}\lrpr{1+z^2}}\frac{1}{\sqrt{\alpha b+1}},
    \end{aligned}
  \end{gathered}
\end{equation}
where $\alpha=2+y^2+z^2$ and $\tld{K}_{12\tau_1\tau_2} = K_{12\tau_1\tau_2} - K_{12\infty\infty}$ as before, $b = \frac{\frac{\ppi}{2} \lrvt{g^2\bar{C}_{12}+I^2\rhoin}}{ 1 +\frac{\ppi}{2} \lrpr{C^{\nam{h}}_{0} -\lrvt{g^2\bar{C}_{12}+I^2\rhoin}}}$, and
$c_{i=1,2} = \sqrt{\frac{1 +\frac{\ppi}{2} \lrpr{C^{\nam{h}}_{0}-C^{\nam{h}}_{\tau_i}}}{ 1 +\frac{\ppi}{2} \lrpr{C^{\nam{h}}_{0}+C^{\nam{h}}_{\tau_i}-2\lrvt{g^2\bar{C}_{12}+I^2\rhoin}}}}$.
Note that unlike in \eqref{eq:2-repl-ch-owen}, \eqref{eq:2-repl-tld-fluct-owen} here contains absolute values.
This allows us to predict $\oOS$ in \eqref{eq:os-tld-result}.

Moving to the level of $\dbl{\bar{C}}$, the overall structure of the Hessian is similar to that in \eqref{eq:2-repl-tld-hess} with the analogous ordering, except every block now becomes a scalar, so the $\XX$-s cannot be approximated away.
One could explicitly invert the entire Hessian since it is only $6\times 6$ with scalar entries, but we note that since the fluctuation is subleading, i.e., $\ep{\bar{C}_{12}}{\dbl{\bar{C}}}\sim 1/\nni$, we would only use it when $\bar{C}_{12}$ itself vanishes.
So when $\bar{C}_{12}=0$ at the saddle-point, i.e., the two external inputs are independent, $\rhoin=0$,
\begin{equation}
  \label{eq:2-repl-bar-hess}
  H = \sbmx{H^{\dbl{\bar{\nam{C}}}_{11}\dbl{\bar{\nam{C}}}_{11}}&0&0
    \\\cdots&H^{\dbl{\bar{\nam{C}}}_{22}\dbl{\bar{\nam{C}}}_{22}}&0
    \\\cdots&\cdots&H^{\dbl{\bar{\nam{C}}}_{12}\dbl{\bar{\nam{C}}}_{12}}},
\end{equation}
and the two single-replica and the cross-replica blocks can again be inverted independently.
The $2\times 2$ cross-replica Hessian can be found to be
\begin{equation}
  \label{eq:2-repl-bar-hess-12}
  \begin{aligned}
    H^{\dbl{\bar{\nam{C}}}_{12}\dbl{\bar{\nam{C}}}_{12}}_{t_1t_2,t'_1t'_2} &= \nni\pmx{0 &-\ii\lrpr{1 +g^2 \nt{\ep{\crc{h}_{1t_1}\crc{h}_{2t_2}, \bar{r}_1\bar{r}_2}{\xi}}{t_1t_2}} \\\cdots &-\ep{\bar{r}_{1}\bar{r}_{2}, \bar{r}_{1}\bar{r}_{2}}{\xi}},
  \end{aligned}
\end{equation}
yielding the fluctuation of $\bar{C}_{12}$ in terms of the single-replica statistics $\bar{C}$:
\begin{equation}
  \label{eq:2-repl-bar-inv-non-non}
  \begin{aligned}
    \ep{\bar{C}_{12}^2}{\dbl{\bar{C}}} =&\frac{1}{\nni}\frac{\bar{C}^2}{\lrpr{1-g^2\ep{r'_t}{\xi}^2}^2}.
  \end{aligned}
\end{equation}
When $\phi\ev{h} = \oerf\ev{\sqrt{\ppi}h/2}$, the expectation can be found to be
\begin{equation}
  \label{eq:2-repl-bar-fluct-owen}
  \ep{r'_t}{\xi}^2 = \frac{1}{1+\frac{\ppi}{2}C^{\nam{h}}_{0}},
\end{equation}
and this allows for numeric predictions in \eqref{eq:pr-bar-result} and \eqref{eq:finite-pr-bar-result}.

\subsection*{Finite sampling quantities}\phantomsection\label{sec:finite}

In this section we obtain the expressions for finite measurement time or number of behavioral contexts.
And we first derive \eqref{eq:finite-pr-bar-result} since the number of behavioral contexts is simpler to deal with compared to measurement time.
For a finite number of contexts $\nnc$, the expression for $\Bar{\oPR}_\nnc$ is
\begin{equation}
  \label{eq:finite-pr-bar}
  \Bar{\oPR}\ev{\bar{\Sigma}_{\nnc,ij}} = \frac{\lrpr{\oTr\ev{\bar{\Sigma}_{\nnc}}}^2}{ \nni\oTr\ev{\bar{\Sigma}\idc{_{\nnc}^2}}} =\frac{\nni^2\lrpr{\frac{1}{\nnc}\sum_a\bar{C}_{aa}}^2}{ \frac{\nni^3}{\nnc^2}\sum_{ab}\bar{C}_{ab}^2},
\end{equation}
and this is nothing more than explicitly writing out the expectation in \eqref{eq:cov-bar-sq}.
When $\rhoin\neq 0$, the saddle-point value of $\bar{C}_{ab}$ is $\sim_\nni 1$ with variance $\sim 1/\nni$, so
\begin{equation}
  \label{eq:corr-cov-bar-sq}
  \bar{C}_{ab}^2=\ep{\bar{C}_{ab}^2}{f_{a,i}f_{b,i}}
\end{equation}
is true to the leading order.
When $\rhoin=0$, the saddle-point value of $\bar{C}_{ab}$ is $0$ with variance $\sim 1/\nni$, and either there is a reasonable number of uncorrelated pairs $f_a,f_b$ sampled by the sum in \eqref{eq:finite-pr-bar} so the equality holds overall, or the number of such terms is small enough so the sum will be dominated by $\bar{C}_{ab}\sim_\nni 1$ when $\rhoin\neq 0$.
The numerator, on the other hand, is unaffected by finite $\nnc$ since the same-replica quantity $\bar{C}_{aa}$ always has non-vanishing self-averaging saddle-point values with subleading fluctuations.
\eqref{eq:finite-pr-bar-result} is then the specific case of $\rhoin=0$ between all pairs of $f$, additionally separating $\bar{C}_{aa}$ from $\bar{C}_{a\neq b}$ in the denominator.

We now move to the effect of finite measurement times, defined by \eqref{eq:weight-1}, \eqref{eq:finite-stat}, and \eqref{eq:finite-cov}.
In general, for two finite time statistics with abbreviations
\begin{equation}
  \label{eq:finite-property-if}
  \begin{aligned}
    &f_{t+\bd{\tau}} = f_{t,t+\tau_1,t+\tau_2,\cdots},
    \qquad&&g_{t'+\bd{\tau}'} = g_{t',t'+\tau'_1,t'+\tau'_2,\cdots},
    \\ &\ep{f_{t+\bd{\tau}}}{t:T,t_\mea}=\int\dd t \,w_{T t_\mea,t}f_{t+\bd{\tau}},
    \qquad&&\ep{g_{t'+\bd{\tau}'}}{t':T,t_\mea}=\int\dd t' \,w_{T t_\mea,t'}g_{t'+\bd{\tau}'},
  \end{aligned}
\end{equation}
they have properties
\begin{equation}
  \label{eq:finite-property-then}
  \begin{gathered}
    \Ep{\ep{f_{t+\bd{\tau}}}{t:T,t_\mea}}{t_\mea}  =\ep{f_{t+\bd{\tau}}}{t},\\
    \begin{aligned}
      \Ep{\ep{f_{t+\bd{\tau}}}{t:T,t_\mea} \ep{g_{t'+\bd{\tau}'}}{t':T,t_\mea}}{t_\mea} &= \int\frac{\dd t\dd t'}{T_\infty} \orelu\Ev{\frac{1-\lrvt{t'-t}/T}{T}} f_{t+\bd{\tau}}g_{t'+\bd{\tau}'}
      \\ &= \int\dd \tau'' \orelu\Ev{\frac{1-\lrvt{\tau''}/T}{T}} \ep{f_{t+\bd{\tau}}g_{t+\tau''+\bd{\tau}'}}{t},
    \end{aligned}
  \end{gathered}
\end{equation}
which can be shown by first performing the outer average over the center of measurement $t_\mea$, referring to \eqref{eq:weight-1}.
Then a simple application of $f_t=g_t=r_{it}-\bar{r}_i$ would lead to \eqref{eq:finite-bar-var}.

To obtain \eqref{eq:finite-pr-tld-result}, we note that the PR \eqref{eq:pr-tld} evaluated at the measured temporal covariance \eqref{eq:finite-cov} averaged over measurements $t_\mea$ is
\begin{equation}
  \label{eq:finite-pr-tld}
  \begin{aligned}
    \ep{\Tld{\oPR}\ev{\tld{\Sigma}_{T t_\mea ij}}}{t_\mea} &= \Ep{\frac{\lrpr{\oTr\ev{\tld{\Sigma}_{T t_\mea ij}}}^2}{\nni\oTr\ev{\tld{\Sigma}\idc{_{T t_\mea ij}^2}}}}{t_\mea}=\Ep{\frac{\nni^2\ep{C_{tt}}{t:T,t_\mea}^2}{\nni^3\ep{\tld{C}_{t_1t_2}^2}{t_1,t_2:T,t_\mea}}}{t_\mea}
    \\ &\approx \frac{\ep{\ep{C_{tt}}{t:T,t_\mea}^2}{t_\mea}}{\nni\ep{\ep{\tld{C}_{t_1t_2}^2}{t_1,t_2:T,t_\mea}}{t_\mea}},
  \end{aligned}
\end{equation}
abbreviated as \eqref{eq:finite-property-if}, where the second line assumes that the denominator has small fluctuations over $t_\mea$.
We expect this to be true for reasons similar to that for \eqref{eq:corr-cov-bar-sq}: if $T$ is small then $\tld{C}_{t_1t_2}\neq 0$ with subleading fluctuations, and if $T$ is large then $\tld{C}_{t_1t_2}=0$ dominates and there would be enough sampling for $\tld{C}_{\infty}^2$.
The numerator can be treated with \eqref{eq:finite-property-then} and $f_{t}=g_{t}= C_{tt}$, and the denominator can be treated by noticing \eqref{eq:finite-property-then} does not use the fact that $f_tg_{t'}$ is a product, so we can do $f_{t}g_{t'}=\tld{C}_{tt'}^2$ to get
\begin{equation}
  \label{eq:finite-cov-tld-sq}
  \begin{aligned}
    \nni\ep{\ep{\tld{C}_{t_1t_2}^2}{t_1,t_2:T,t_\mea}}{t_\mea} = &\nni\int\dd \tau \orelu\Ev{\frac{1-\lrvt{\tau}/T}{T}} \ep{\tld{C}_{t,t+\tau}^2}{t}
    \\=&\nni\Lrpr{\ep{\tld{C}_{t,t+\infty}^2}{t}+\int\dd \tau \orelu\Ev{\frac{1-\lrvt{\tau}/T}{T}} \tld{C}_{\tau}^2},
  \end{aligned}
\end{equation}
leading to \eqref{eq:finite-pr-tld-result}.
The second line of \eqref{eq:finite-cov-tld-sq} follows from separating the fluctuation of $\tld{C}_\tau$ from its saddle-point value and noting that the fluctuation at $\tau\sim\tau_\acov$ ($\tld{C}_{\tau\neq\infty}$) does not contribute to leading order for both $\tau_\acov\sim T$ and $\tau_\acov\ll T$.
Alternatively, \eqref{eq:finite-cov-tld-sq} could be obtained from the two-site cavity view~\cite{clark2023Dimension}, where we apply \eqref{eq:finite-property-then} to the measured covariance $\tld{\Sigma}_{T t_\mea ij}$ directly, separating the same-neuron terms from the cross-neuron terms.

Finally, we derive \eqref{eq:finite-os-result}.
According to \eqref{eq:os}, the OS at finite times is
\begin{equation}
  \label{eq:finite-os-tld}
  \oOS_{T} =\Ep{\frac{\oTr\ev{\tld{\Sigma}_{T t_{\mea 1} 1}\tld{\Sigma}_{T t_{\mea 2} 2}}}{ \sqrt{\oTr\ev{\tensor{\tld{\Sigma}}{_{T t_{\mea 1} 1}^2}}\oTr\ev{\tensor{\tld{\Sigma}}{_{T t_{\mea 2} 2}^2}}}}}{t_{\mea 1}t_{\mea 2}} \approx\frac{\ep{\oTr\ev{\tld{\Sigma}_{T t_{\mea 1} 1}\tld{\Sigma}_{T t_{\mea 2} 2}}}{t_{\mea 1}t_{\mea 2}}}{\nni\ep{\ep{\tld{C}_{t_1t_2}^2}{t_1,t_2:T,t_\mea}}{t_\mea}}.
\end{equation}
Here, as in the case for \eqref{eq:os-tld-result}, the two replicas are statistically identical with independent thermal fluctuations.
Consequently, the denominator is the same as that in \eqref{eq:finite-pr-tld}, so we again assume its fluctuations are small over $t_\mea$, moving the outer expectation to the numerator.
Then, using \eqref{eq:finite-property-then} with $f_{t+\bd{\tau}} = \tld{r}_{it}\tld{r}_{j,t+\tau}$, we see that \eqref{eq:finite-os-tld} has the same numerator as \eqref{eq:os-tld-result}.
And comparing them to \eqref{eq:pr-tld-result} and \eqref{eq:finite-pr-tld-result}, we get \eqref{eq:finite-os-result}.

\subsection*{Numerics}\phantomsection\label{sec:numerics}

\subsubsection*{Network simulation}

\paragraph*{Model and integration}
Recall that the network in \eqref{eq:ode} is simulated with $J_{ij}\fl\oNor\ev{0,g^2/\nni}$ and time-independent external inputs $f_i\fl\oNor\ev{0,I^2}$.
We integrate the dynamics with a fourth-order Runge--Kutta method using time step $\Delta t=0.1$, and use the same nonlinearity as in the analysis, $\phi\ev{h}=\oerf\ev{\sqrt{\ppi}h/2}$.
Unless otherwise stated, simulations in the main text use $\nni=800$, $g=3$, and $I\in\lrbe{0,0.9,1.8,2.7,3.6}$.
Initial conditions are sampled independently from a standard Gaussian distribution.

\paragraph*{Trajectory statistics}
For each simulated trajectory over time, we discard an initial transient of duration $50$ and then collect statistics over a stationary window of duration $24000(=30\nni)$.
We choose this long window because the theory predicts that the true dimensionality converges slowly in measurement time $T$ in units of $\nni$.
Long-time quantities in the main text are approximated using quantities with $T=24000$.
In particular, we compute $\bar{C}$, $\tld{C}_0$, $\tld{C}_\tau$, $\Tld{\oPR}_\infty$, $\oCS$, and $\oOS$ according to their definitions in the main text.
Note that all simulated covariances are computed without an additional finite-time mean subtraction, as defined and justified in the main text.

\paragraph*{One- and two-context quantities}
For the single-context and two-context quantities, we average over $200$ independent realizations for each parameter set, where each realization consists of one network $J_{ij}$ together with the sampled external inputs.
For the similarity curves, each network is paired with $6$ correlated external inputs with prescribed pairwise similarities $\rhoin=\cos(n\pi/10)$ for $n=0,1,\cdots,5$.
This lets one batch of simulations provide all prescribed input similarities.
Compared to generating each pair separately, it mainly makes the samples less independent, but it reduces the total computation substantially.
In all figures, markers show the mean over realizations and error bars show the standard deviation across realizations.

\paragraph*{Multi-context quantities}
For the multi-context quantities, we average over $40$ independent realizations for each parameter set.
For each realization and each nonzero external input strength $I\in\lrbe{0.9,1.8,2.7,3.6}$, we sample $8000(=10\nni)$ independent external inputs and use the corresponding ordered responses to estimate the multi-context quantities.
Since these quantities depend on the cloud of ordered responses rather than on dynamic geometry, each ordered response is estimated from a shorter stationary window of duration $250$ after the transient of duration $50$.
This shorter window is part of the computational cost tradeoff, and is justified because the error in the ordered response decreases on the scale of $\tau_{\tld C}$, which is small compared to the timescale needed for the true dimensionality to converge.
We use fewer realizations here because each multi-replica quantity requires much more computation than the other quantities, so it is impractical to use the same total number of realizations for both.
This is also why the numerical multi-context dimensionality values are slightly lower than the predicted ones when the dimensionality is high: the high-dimensional cloud of ordered responses is hard to sample uniformly.
At the same time, the multi-context dimensionality is still well predicted and strongly self-averaging, with very small standard deviations across realizations.
From the same sampled ordered responses, we also compute the fluctuation-dissipation quantity in \apperef{sec:fluct-dissp}, namely the $\oOS$ between the covariance of ordered responses over contexts and the autonomous temporal-chaos covariance.

\paragraph*{Finite-time and finite-context sampling}
For the finite-time and finite-context calculations, we do not average over all possible windows or subsets.
Instead, for each window length or context number, we use at most $20$ folded samples.
This is a practical restriction, since each sample requires forming and analyzing an $\nni\times\nni$ covariance matrix.
This is a conservative numerical limitation: it can only make the numerical agreement with theory look worse, not better.

\subsubsection*{Semi-analytic numerics}\phantomsection\label{sec:ana-numerics}

\paragraph*{Numerical solution of the one-replica theory}
All semi-analytic predictions begin by numerically solving the one-replica DMFT equation for the stationary temporal-chaos autocovariance $\tld{C}_\tau$, using the effective-potential formulation introduced in Section~\nameref{sec:dmft1-sp}.
Finding $\tld{C}_\tau$ amounts to first identifying the correct initial condition $\tld{C}_0$ and solving the differential equation \eqref{eq:1-repl-ch-owen}.

As in previous studies, the relevant solution in the chaotic regime is the stable decaying one, and this specifies the correct initial condition~\cite{crisanti2018Path}.
For each pair of parameters $(g,I)$, we therefore use an outer loop over candidate values of the zero-lag autocovariance $\tld{C}_0$, and for each candidate construct the corresponding effective force and potential to determine the self-consistent solution.
The inner step amounts to locating the zero-force point, equivalently the maximum of the potential, for that candidate value of $\tld{C}_0$.
We then refine the outer search by repeated zoom-in and interpolation to determine $\tld{C}_0$ with high precision, and finally evaluate the full function $\tld{C}_\tau$ on a finite grid in $\tau$.
In practice, the decaying solution is numerically sensitive to the value of $\tld{C}_0$: if the inferred initial potential is slightly too large, the trajectory crosses the potential barrier and gives the wrong solution, whereas if it is slightly too small, the trajectory develops oscillations.
When the mismatch between the two endpoints of the potential is sufficiently small, we linearly interpolate this residual difference to regularize the potential so that the endpoints agree.

Given the numerically identified initial condition $\tld{C}_0$, we then integrate \eqref{eq:1-repl-ch-owen} using a standard numerical ODE solver, such as those provided in \texttt{scipy}.
In general, $\tld{C}_\tau$ is localized near $\tau=0$ with width $\tau_\acov$.
Networks considered in this work are typically away from criticality, where $\tau_\acov$ is of order $5$.
So we truncate the simulation at $\tau = 100$, corresponding to about $20\tau_\acov$.

\paragraph*{Evaluation of theory quantities from $\tld{C}_\tau$}
Once $\tld{C}_\tau$ is obtained numerically, the fluctuation quantities around the one-replica and two-replica saddle points are evaluated from $\tld{C}_\tau$ through the Fourier-space relations derived in Section~\nameref{sec:dmft1-fluct} and Section~\nameref{sec:dmft2}.
Since discrete Fourier transforms assume periodic inputs, we mirror $\tld{C}_\tau$ to negative lags before transforming to reduce edge artifacts.

We then evaluate all remaining theory quantities, including $\tld{C}_0$, $\Tld{\oPR}_\infty$ and its finite-$T$ correction, $\oCS$, $\oOS$, and  $\Bar{\oPR}_\infty$ and its finite-$\nnc$ correction, by numerically applying the formulas given in the main text and Methods.
The calculations for long-time quantities are only algebraic.
The finite time quantities additionally require numerical quadrature, in which case interpolation is done when it improves computational efficiency.

\paragraph*{Parameter grids and conventions}
Theory curves are evaluated on parameter grids denser than the corresponding simulation points in order to draw smooth semi-analytic curves as functions of $I$, $\rhoin$, or $T/\nni$.
For the robustness calculations in \apperef{sec:g-indp}, the search window used to solve the one-replica self-consistency problem is enlarged when necessary at large $g$.

\subsubsection*{Experimental data extraction}

\paragraph*{Plot digitization}
For the experimental comparisons in Figures~\ref{fig:finite-tld} and~\ref{fig:bar}, we digitized the faint red data points from the corresponding published plots using WebPlotDigitizer \url{https://apps.automeris.io/wpd4/}.
Specifically, the data for Figure~\ref{fig:finite-tld} were extracted from Figure~4C of \citep{gao2017Theory}, and those for Figure~\ref{fig:bar} were extracted from Figure~3G,H of \citep{bartolo2020Dimensionality}.
In each case, the axes were calibrated to the published plot, and the data points were digitized as $(x,y)$ coordinate pairs, with the vertical coordinate used directly as the reported dimensionality.

\paragraph*{Processing for temporal-chaos dimensionality data}
For Figure~\ref{fig:finite-tld}, the digitized horizontal coordinate gives the measurement-time axis reported by \citep{gao2017Theory}, which we rescale to match the autocorrelation-time convention used in this paper.
In \citep{gao2017Theory}, the autocorrelation time is the lag at which a Gaussian fit to the stationary autocorrelation decreases to $1/\ee$ of its zero-lag value, whereas our $\tau_\acov$ in \eqref{eq:tau-acov} is defined as the integral of the squared normalized autocovariance.
For a Gaussian autocovariance, converting from the $1/\ee$ decay lag to our convention requires multiplying the digitized time axis by $\sqrt{2/\ppi}$.
We leave the dimensionality values unchanged.
For visualization, we pooled the digitized and rescaled points across the extracted data series and divided them into bins along the rescaled measurement-time axis.
Bins containing only one point were merged with neighboring sparse bins, and the plotted experimental summary shows the mean and standard deviation of both coordinates within each resulting bin.

\paragraph*{Processing for multi-context dimensionality data}
For Figure~\ref{fig:bar}, the digitized Bartolo data consist of four value series corresponding to four experimental conditions.
We converted the block number reported in \citep{bartolo2020Dimensionality} to the number of behavioral contexts by multiplying by two, because each block contains two stimuli.
The faint red error bars in Figure~\ref{fig:bar} show the individual digitized condition series with their reported errors.
The visible red data series summarizes the four conditions by plotting their mean dimensionality at each context number.
Its error bar represents the spread of the dimensionality values when all four conditions are pooled together, including the reported error within each condition.


\appendix
\section*{Generality of non-monotonicity in dimensionality of temporal chaos}\phantomsection\label{sec:g-indp}

\begin{figure}
  \centering
  \includegraphics{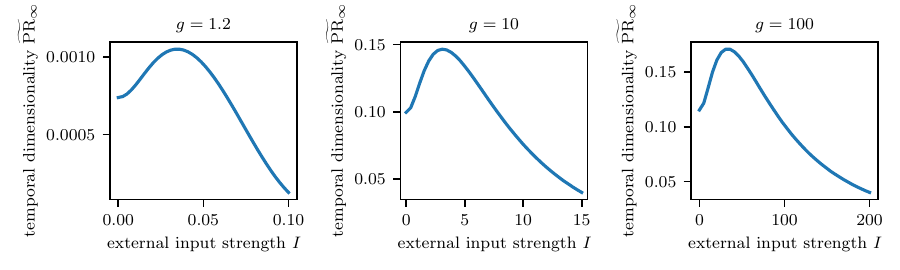}
  \caption{The dimensionality $\Tld{\oPR}_\infty$ is non-monotonic in external input strength $I$ over orders of magnitude of the gain parameter $g$.
    The range of external input strength $I$ changes with $g$ because the transition to ordered dynamics varies with $g$.
  }
  \label{fig:app-1}
\end{figure}

To verify that the non-monotonic dependence of the long-time temporal dimensionality $\Tld{\oPR}_\infty$ on the external input strength $I$ is not specific to the parameter choice used in the main text, we evaluate the semi-analytic prediction for gain values $g=1.2$, $10$, and $100$, spanning nearly two orders of magnitude.
Figure~\ref{fig:app-1} shows that in all three cases, $\Tld{\oPR}_\infty$ increases at small $I$, reaches a maximum at intermediate input strength, and then decreases at larger $I$.
Thus, although the location and scale of the peak vary with $g$, the qualitative non-monotonic dependence on input strength is robust across gain values spanning orders of magnitude.

\section*{Susceptibility of time-averaged response}\phantomsection\label{sec:fluct-dissp}

\begin{figure}[t!]
  \centering
  \includegraphics{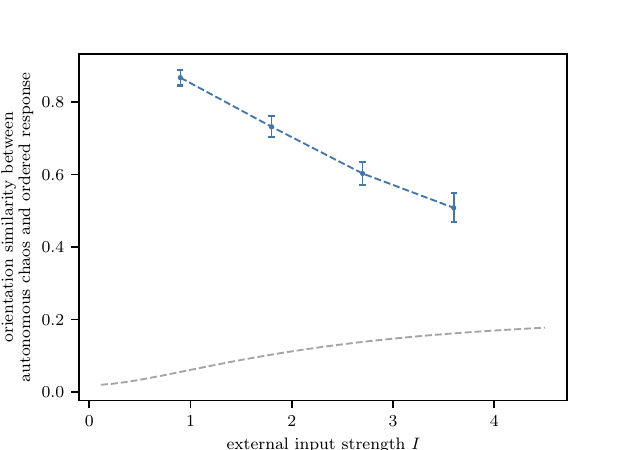}
  \caption{The similarity between the orientation of autonomous ($I=0$) temporal chaos and the orientation of ordered responses over behavioral contexts is high for weak external input strengths $I\ll 1$.
    Even though the value decreases greatly as $I$ increases, the similarity is still much higher compared to the value $\Tld{\oPR}_\infty$ expected for uncorrelated orientations shown by the dashed line.
  }
  \label{fig:app-2}
\end{figure}

To test the fluctuation-dissipation interpretation for the initial increase of temporal dimensionality under weak input, we compare the orientation of ordered responses across external inputs with the orientation of autonomous temporal chaos.
Specifically, fixing the coupling matrix $J$, we quantify using $\oOS$ in \eqref{eq:os} the similarity between the orientation of ordered responses over external inputs of strength $I$, represented by $\Bar{\Sigma}$, and the orientation of the autonomous $I=0$ temporal chaos, represented by $\Tld{\Sigma}_\infty$.
As shown in Figure~\ref{fig:app-2}, this similarity is largest at weak input and decreases as the input strength increases, indicating that weak ordered responses are preferentially aligned with the dominant fluctuation directions of the autonomous chaotic state.
This supports the interpretation that weak input initially amplifies activity along pre-existing temporal-chaos directions, thereby contributing to the rise of $\Tld{\oPR}_\infty$.

\section*{Error of finite-time statistics of temporal chaos}\phantomsection\label{sec:error-over-time}

In this section, we show that the error in the finite-time ordered response has variance $\sim\tau_\acov/T$, and we plot the finite-time dimensionality's self-averaging error $\sim \nni/T$.

\begin{figure}[t!]
  \centering
  \includegraphics{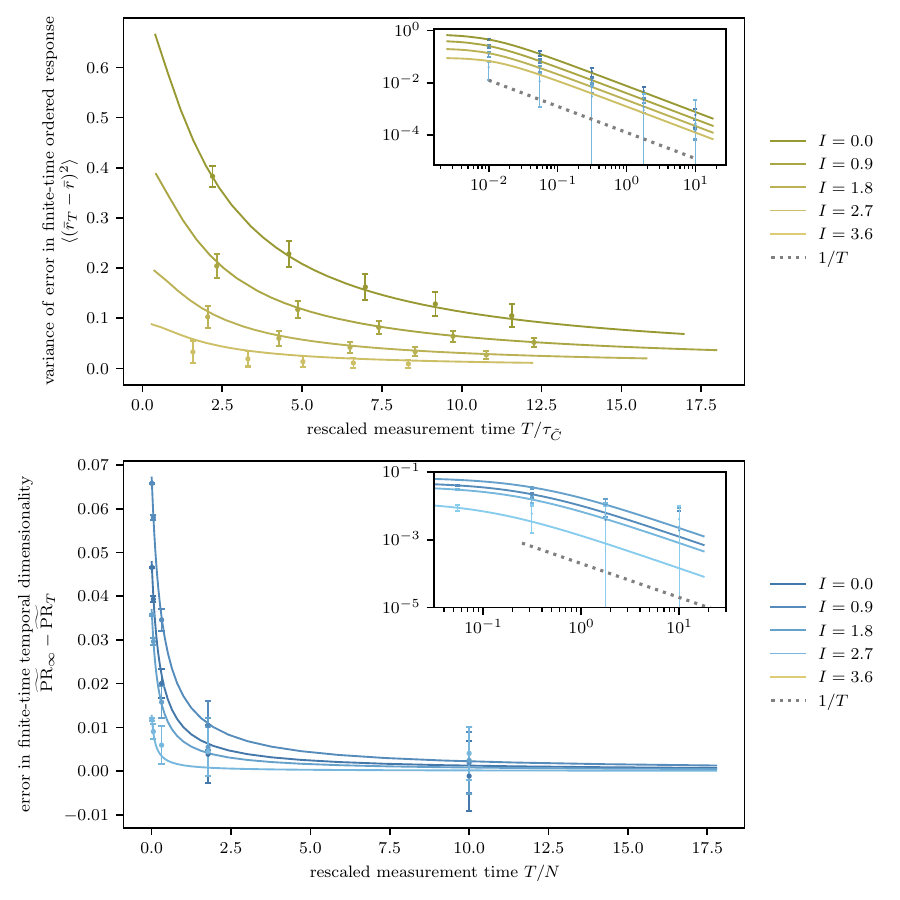}
  \caption{A: The variance of the error in the measured ordered response is $\sim \tau_\acov/T$.
    The inset shows the log-log scale.
    B: The self-averaging error in the finite-time dimensionality is $\sim \nni/T$.}
  \label{fig:error-over-time}
\end{figure}

For the error in the ordered response, since the system is statistically self-averaging and stationary, the variance of the error in the finite-time ordered response $\bar{r}_{T t_\mea}-\bar{r}$ measured for a fixed time window of length $T$ over realizations is equal to its variance over window locations $t_\mea$.
Using results from Section~\nameref{sec:finite}, the resulting expression is
\begin{equation}
  \label{eq:finite-bar-var}
  \Ep{\frac{1}{\nni}\sm{\lrpr{\bar{r}_{T t_\mea i} - \bar{r}_i}^2}{i}}{t_\mea} = \int  \dd \tau\frac{\orelu\ev{1 - \lrvt{\tau}/T}}{T}\tld{C}_\tau,
\end{equation}
which has a very similar form to the integral correction in \eqref{eq:finite-pr-tld-result}.
When the time window is small, $T\ll\tau_\acov$, the $\orelu$ factor varies slowly over the width of $\tld{C}_\tau$, so the integral is dominated by $\tld{C}_0$ and is therefore approximately independent of $T$.
Accordingly, the small-$T$ plateau of the variance inherits the dependence of $\tld{C}_0$ on the external input strength $I$, which decreases with $I$ as shown in Figure~\ref{fig:tld}B.
When $T\gtrsim\tau_\acov$, the integral instead scales as $\sim \tau_\acov/T$, reflecting averaging over approximately $T/\tau_\acov$ effectively independent temporal samples, as expected from the central limit theorem.
These two regimes are shown by the numerical curves in Figure~\ref{fig:error-over-time}A and its log-log inset.
Returning to the justification of \eqref{eq:finite-stat}, in the experimentally relevant regime $T/\tau_\acov \approx 10$, the variance of the finite-time ordered-response error is already small, so replacing $\bar{r}_{T t_\mea}$ by $\bar{r}$ is justified.

We next consider the self-averaging error of the finite-time dimensionality itself.
Unlike the finite-time ordered response, $\Tld{\oPR}_T$ depends on the full $\nni\times\nni$ finite-time covariance matrix of temporal chaos, and therefore converges more slowly with the observation time.
As shown in Section~\nameref{sec:finite}, the leading finite-time correction scales as a power law in the ratio $\nni/T$, as given in \eqref{eq:finite-pr-tld-result-approx}.
Figure~\ref{fig:error-over-time}B confirms this scaling numerically: the self-averaging error of the finite-time dimensionality is controlled by $\nni/T$ and remains substantial even when the error in the finite-time ordered response is already small.
Thus, while the finite-time mean response rapidly approaches its infinite-time limit once $T\gtrsim\tau_\acov$, the finite-time dimensionality converges only slowly, on the scale of $\nni$.

\section*{Similarity between ordered responses after transition to ordered dynamics}\phantomsection\label{sec:cs-sat}

\begin{figure}[t!]
  \centering
  \includegraphics{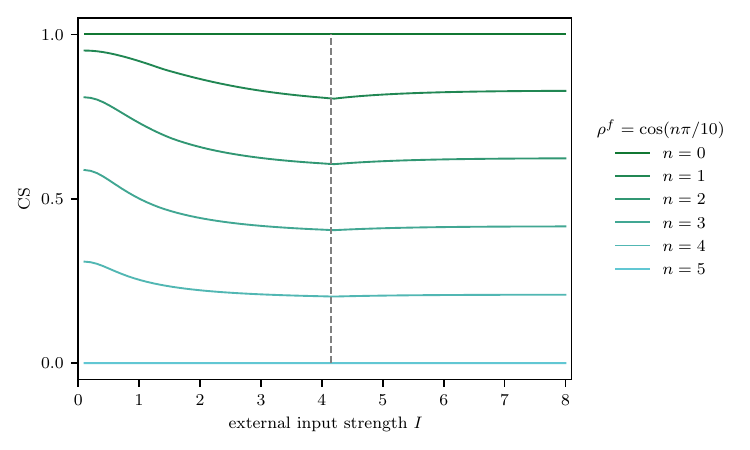}
  \caption{The dependence of the cosine similarity $\oCS$ on the external input strength $I$ over a greater range of $I$ past the transition.
    Curves are labeled by $\rho^f=\cos(n\pi/10)$.
    The decrease in $\oCS$ is local and weak.}
  \label{fig:cs-saturation}
\end{figure}

To confirm that the decrease in the cosine similarity $\oCS$ between the two ordered responses is mainly a local effect near the transition to ordered dynamics, we evaluate $\oCS$ over a wider range of the external input strength $I$ than shown in the main text for input similarities $\rho^f=\cos(n\pi/10)$.
Figure~\ref{fig:cs-saturation} shows that $\oCS$ decreases as the transition is approached, reaches a minimum near the transition, and then varies only weakly at larger $I$.
Thus, while increasing input strength reduces the similarity between the two ordered responses near the transition, this reduction remains limited and does not continue to grow substantially deeper in the ordered regime.

\end{document}